 \documentclass[journal, twocolumn,twoside]{IEEEtran}

\normalsize

\usepackage{cite}
\ifCLASSINFOpdf
   \usepackage[pdftex]{graphicx}
\else
  \usepackage[dvips]{graphicx}
\fi

\usepackage[cmex10]{amsmath}
\usepackage{amssymb}

\usepackage{stfloats}
\newcommand{\upmapsto}{\circlearrowleft}
\newcommand{\Z}{\mathbb{Z}}
\newcommand{\Ftwo}{\mathbb{F}_2} 
\newcommand{\code}[1]{\mathcal{#1}}
\newcommand{\matr}[1]{\mathbf{#1}}
\newcommand{\GF}[1]{\mathbb{F}_{#1}}
\newcommand{\vc}{\vect{c}}
\newcommand{\tr}{\mathsf{T}}
\newcommand{\vect}[1]{\mathbf{#1}}

\newcommand{\dmins}{d_{\mathrm{min}}}
\newcommand{\defeq}{\triangleq}
\newcommand{\cC}{\mathcal{C}}

\def\diag{{\rm diag}}

\newcommand{\base}{\mathbf{B}}
\newcommand{\bpre}{\mathbf{B}^{\uparrow m}}
\newcommand{\bcirc}{\mathbf{B}^{\upmapsto N}}
\newcommand{\hpre}{\mathbf{B}^{\uparrow m \upmapsto r}}

\usepackage{theorem} 

 \newtheorem{theorem}{Theorem}
 \newtheorem{lemma}[theorem]{Lemma} 
  \newtheorem{corollary}[theorem]{Corollary} 
  \newtheorem{remark}[theorem]{Remark} 
\newcounter{mytempeqcounter}

\newcommand{\bigformulatop}[2]{%
  \begin{figure*}[!t]
    \normalsize
    \setcounter{mytempeqcounter}{\value{equation}}
    \setcounter{equation}{#1}
    #2

    \setcounter{equation}{\value{mytempeqcounter}}
    \hrulefill
    \vspace*{4pt}
  \end{figure*}
}

\newcounter{example}
\newenvironment{example}[1][]{\refstepcounter{example}\par\medskip\noindent%
   \textbf{Example~\theexample. #1} \rmfamily}{\medskip}
   
   \newcounter{propertycounter}
\newenvironment{propertyen}[1][]{\refstepcounter{propertycounter}\par\medskip\noindent%
   \textbf{Property~\thepropertycounter. #1} \rmfamily}{\medskip}

\usepackage{color,soul}

\definecolor{lightblue}{rgb}{.90,.95,1}
\sethlcolor{yellow}

\begin{document}
\title{Quasi-Cyclic LDPC Codes \\ based   on Pre-Lifted Protographs}

\author{David~G.~M.~Mitchell,~\IEEEmembership{Member,~IEEE,}
        Roxana~Smarandache,~\IEEEmembership{Member,~IEEE,}
        and~Daniel~J.~Costello,~Jr.,~\IEEEmembership{Life~Fellow,~IEEE}% <-this % stops a space
\thanks{This work was supported in part by the National Science Foundation under Grant Numbers CCF-$1161754$, CCF-$1252788$, and DMS-$1313221$. The material in this paper was presented in part at the IEEE Information Theory Workshop, Paraty, Brazil, October 2011, and in part at the IEEE Information Theory Workshop, Lausanne, Switzerland, September 2012.}% <-this % stops a space
\thanks{D.~G.~M.~Mitchell,  R.~Smarandache, and D.~J.~Costello,~Jr. are with the Department
of Electrical Engineering, University of Notre Dame, Notre Dame,
IN 46556, USA (e-mail: david.mitchell@nd.edu;~rsmarand@nd.edu;~costello.2@nd.edu). D.~G.~M.~Mitchell and  R.~Smarandache are also with the Department of Mathematics, University of Notre Dame, Notre Dame,
IN 46556, USA.}% <-this % stops a space
\thanks{Copyright (c) 2014 IEEE. Personal use of this material is permitted.  However, permission to use this material for any other purposes must be obtained from the IEEE by sending a request to pubs-permissions@ieee.org.}}%
%\thanks{Manuscript received April 19, 2005; revised January 11, 2007.}}

% and the end of the author line. \emph{i.e.}, if you had this:

\markboth{IEEE Transactions Information Theory (Submitted paper)}%
{Submitted paper}



\maketitle

\begin{abstract}

Quasi-cyclic low-density parity-check (QC-LDPC) codes based on protographs are of great interest to code designers because analysis and implementation are facilitated by the protograph structure and the use of circulant permutation matrices for protograph lifting. However, these restrictions impose undesirable fixed upper limits on important code parameters, such as minimum distance and girth. In this paper, we consider an approach to constructing  QC-LDPC codes that uses a two-step lifting procedure based on a protograph, and, by following this method instead of the usual one-step procedure, we obtain improved minimum distance and girth properties. We also present two new design rules for constructing good QC-LDPC codes using this two-step lifting procedure, and in each case we obtain a significant increase in minimum distance  and achieve a certain guaranteed girth compared to one-step circulant-based liftings. The expected performance improvement is verified by simulation results.
\end{abstract}

\begin{IEEEkeywords}
Low-density parity-check (LDPC) codes, girth, minimum distance, protograph, quasi-cyclic codes, Tanner graph.
\end{IEEEkeywords}

% \fi
\IEEEpeerreviewmaketitle

\section{Introduction}\label{sec:intro}
\IEEEPARstart{A}{} \emph{protograph} \cite{tho03} is a small Tanner graph \cite{tan81} described by an $n_c \times n_v$ biadjacency matrix $\matr{B}$, known as a base matrix, that consists of non-negative integers $B_{i,j}$ that correspond to $B_{i,j}$ parallel edges in the graph. A protograph-based code is obtained by taking an $N$-fold graph cover \cite{mas77}, or ``lifting'', of a given protograph and can be described by an $Nn_c\times Nn_v$ parity-check matrix obtained by replacing each non-zero entry $B_{i,j}$  by a sum of $ B_{i,j}$ non-overlapping   permutation matrices of size $N\times N$ and each zero entry by an $N\times N$ all-zero matrix. The set of all such codes that can be derived from the protograph in this fashion is referred to as a code \emph{ensemble}. Low-density parity-check (LDPC) code ensembles \cite{gal62} based on a protograph form a subclass of multi-edge type codes \cite{ru02} that, for suitably-designed protographs, have many desirable features, such as good iterative decoding thresholds and linear minimum distance growth, \emph{i.e.}, they are asymptotically good (see, \emph{e.g.}, \cite{ddja09,ady07,mpc13}).

The construction of \emph{quasi-cyclic} LDPC (QC-LDPC) codes \cite{klf01,fos04, txla05, myk05, myk06,lcz+06,szla09,khz+10,zhl+10,zla+11,cxdl04,tss+04,sv12,bhj+12} can be seen as a special case of the protograph-based construction in which the $N$-fold cover is obtained by restricting the edge permutations to be cyclic and can be described by an $Nn_c \times Nn_v$  parity-check matrix formed as an $n_c \times n_v$ array of $N\times N$ circulant matrices. Members of a protograph-based LDPC code ensemble that are QC are particularly attractive from an implementation standpoint, since they can be encoded with low complexity using simple feedback shift-registers \cite{myk05,lcz+06} and their structure leads to efficiencies in decoder design \cite{wc07b,dyc09}. Moreover, QC-LDPC codes can be shown to perform well compared to random LDPC codes for moderate block lengths \cite{klf01,cxdl04,tss+04,khz+10}.  However, unlike typical members of an asymptotically good protograph-based LDPC code ensemble, the QC sub-ensemble does not have linear distance growth. Indeed, if the protograph base matrix consists of only ones and zeros, then the minimum Hamming distance is bounded above by $(n_c+1)!$, where $n_c$ is the number of check nodes in the protograph, regardless of the lifting factor $N$ \cite{md01,fos04}. 

A great deal of research effort has been devoted to designing QC-LDPC codes with large girth (see, \emph{e.g.}, \cite{fos04,osu06,myk06,bhj+12}) and minimum distance (see, \emph{e.g.}, \cite{klf01,fos04,khz+10,zhl+10,zla+11,cxdl04,tss+04,sv12}). QC-LDPC codes based on protographs have also been designed to improve certain characteristics, such as girth \cite{kncs07,bch08,phns13} or lowering the ``error floor'' \cite{aba11}.  In \cite{kb12}, lower bounds on the size of the necessary lifting factor  $N$ of a protograph required to achieve a  certain girth is investigated for QC-LDPC codes derived from several simple protograph types. Minimum distance bounds for protograph-based QC-LDPC codes were presented in \cite{sv12} and later improved for several cases in \cite{bs13}. {A useful feature of the approach presented in the current paper, which we will demonstrate later, is that a good existing QC-LDPC code design can be used in conjunction with our methodology to improve code performance.}

Several authors have also considered two- or multi-step liftings of a base graph. In \cite{ccsds07}, irregular protograph-based QC-LDPC codes are proposed with parallel edges in the protograph. In order to have only single edges in the code's Tanner graph, which is desirable for efficient implementation, the authors first employ an $m$-step expansion, where $m$ is sufficiently large to disperse the parallel edges, before applying a second lifting step. In \cite{sv12}, the authors present an example showing that a QC code obtained from a double lifting has a larger minimum distance than a single lifted code. Also, so-called hierarchical QC-LDPC codes have been constructed in a recent paper \cite{wdy13}. These {high-girth} constructions are obtained by taking repeated circulant-based liftings of a base graph, but the authors do not consider minimum distance. %Here, we take a joint approach to constructing protograph-based QC-LDPC codes with both large girth and large minimum distance. 

In this paper, we investigate QC-LDPC codes that are constructed using a two-step lifting procedure based on a protograph: a ``pre-lifting'' step where we take an $m$-fold graph cover of the protograph, where $m$ is typically small, and a ``second-lifting'' step where we take an $r$-fold graph cover of the \emph{pre-lifted} protograph, where $r$ is typically large and the permutations are chosen to be cyclic.\footnote{It is also possible to construct QC-LDPC codes in this way using more than two lifting steps, but only two-step liftings are considered in this paper.} As a result of the pre-lifting, we can construct QC-LDPC codes with increased girth \emph{and} minimum distance while maintaining the circulant-based structure that facilitates efficient implementation. In particular, we show that the QC-LDPC code ensemble obtained from a pre-lifted protograph can have an increased upper bound on minimum distance compared to the QC-LDPC code ensemble obtained from the original protograph and we demonstrate the existence of codes with minimum distance exceeding the original bound. We also present two design rules for code construction: one uses only commuting pairs of permutation matrices at the first (pre-lifting) stage, while the other uses at least one pair of non-commuting permutation matrices. In each case, we obtain a significant increase in the minimum distance  and achieve a certain guaranteed girth compared to a one-step circulant-based lifting of the original protograph. The expected performance improvement is verified by simulation results.

The paper is structured as follows. In Section \ref{sec:background}, we provide the necessary background material, describe the structure of the QC sub-ensemble of a protograph-based LDPC code ensemble, and review existing bounds concerning the minimum Hamming distance of QC-LDPC protograph-based codes. In Section \ref{sec:prelifting}, we introduce the concept of pre-lifting, discuss some necessary conditions to permit increased minimum distance and girth for our construction technique, and present two new code design rules. In Section \ref{sec:constructingprelifted}, we focus on the pre-lifting step and derive circulant-based codes with minimum distance and girth exceeding the original bounds for QC codes without pre-lifting. Sections~\ref{sec:liftcirculant} and \ref{sec:liftnoncommuting} demonstrate the application of the two code design rules. The expected performance improvement is verified by simulation results. In Section \ref{sec:bocharova} we construct a nested family of QC-LDPC codes with design rates $R=1/4, 2/5, 1/2$, and $4/7$ and robustly good performance by applying the pre-lifting technique to a QC-LDPC code with large girth taken from the literature. Finally, concluding remarks are given in Section~\ref{sec:conclusions}.

%%%%%%%%%%%%%%%%%%%%%%%%%%%%%%%%%%%%%%%%%%%%%%%%%%%
%%%%%%%%%%%%%%%%%%%%%%%%%%%%%%%%%%%%%%%%%%%%%%%%%%%
\section{Basic definitions, notation,  and background} \label{sec:background}

%%%%%%%%%%%%%%%%%%%%%%%%%%%%%%%%%%%%%%%%%%%%%%%%%%%
%%%%%%%%%%%%%%%%%%%%%%%%%%%%%%%%%%%%%%%%%%%%%%%%%%%
\noindent {\it Notation}  %We use the following sets:\\
\begin{itemize} 
\item For any positive integer $L$,
$[L]\defeq\{ 0, 1, \ldots, L{-}1 \}$.
\item $\Z $ is the ring of integers;
 $\Ftwo$ is the Galois field of size $2$. 
\item $\Ftwo^n$ and  $\Ftwo^{k \times n}$ are, respectively, the set of row vectors over $\Ftwo$ of length $n$
and the set of matrices over $\Ftwo$ of size $k \times
n$. 
\end{itemize}

{\it Linear codes}
\begin{itemize}\item All the codes in this paper are binary linear codes. \item An $[n,k,\dmins]$ {\em linear code} $\cC$ of length $n$, dimension $k$, and minimum Hamming distance $\dmins$ can be specified as the null space of a $p \times n$ (scalar) parity-check matrix $\matr{H} \in \GF{2}^{p \times n}$, where the rank of the matrix is $n-k\leq p$, \emph{i.e.}, $$\code{C} = \bigl\{ \vc \in
\Ftwo^n \ \bigl| \ \matr{H} \cdot \vc^\tr = \vect{0}^\tr \bigr. \bigr\},$$
where ${}^\tr$ denotes transposition.
%\item The {\em minimum distance} associated with a particular code or the null space of a parity-check matrix $\matr{H}$ will be denoted by  $\dmins$. 
\end{itemize}
\noindent{\it Tanner graphs}
\begin{itemize}
\item With a parity-check matrix $\matr{H}$ we associate a bipartite {\em Tanner
graph}~\cite{tan81} in the usual way.  
\item The {\em girth} of a Tanner graph associated with a parity-check matrix $\matr{H}$ is the length of the shortest cycle in the graph and is denoted by $g$.
%\end{itemize}

\end{itemize}
%\subsection{Graphs}
%\label{sec:notation:1:graphs}
%XXXX Change this part. 
%***************************************************************************
%
%With a parity-check matrix $\matr{H}$ we associate a Tanner
%graph~\cite{tan81} in the usual way.
%namely, for every code bit we draw a
%variable node, for every parity-check we draw a check node, and we connect a
%variable node and a check node by an edge if and only if the corresponding
%entry in $\matr{H}$ is nonzero. Similarly, the Tanner graph associated to a
%polynomial parity-check matrix $\matr{H}(x)$ is simply the Tanner graph
%associated to the corresponding (scalar) parity-check matrix $\matr{H}$.
%As usual, the degree of a vertex is the number of edges incident to it and an
%LDPC code is called $(d_1, d_2)$-regular if all variable nodes have degree
%$d_1$ and all check nodes have degree $d_2$. Otherwise,  the
%code is {\em irregular}. 
%A simple cycle of a graph will be a
%backtrackless, tailless, closed walk in the graph, and the length of such a
%cycle is defined to be equal to the number of visited vertices (or,
%equivalently, the number of visited edges). 
%The girth of a graph is then the
%length of the shortest %simple 
%cycle in  the graph.
%%%%%%%%%%%%%%%%%%%%%%%%%%%%%%%%%%%%%%%%%%%%%%%%%%%
%\subsection{Permutations and permutation matrices}\label{sec:permutations}
%%%%%%%%%%%%%%%%%%%%%%%%%%%%%%%%%%%%%%%%%%%%%%%%%%%
{\it Permutations} 
\begin{itemize}\item An $N$-{\em permutation} $\sigma$ is a one-to-one function on the set $\mathcal{N}=\{1,2,\ldots,N\}$ described as:
\[\sigma\defeq \left(\begin{matrix}
       1 & 2 &\cdots & N\\
       \sigma(1) &  \sigma(2) & \cdots & \sigma(N)
     \end{matrix}\right).
\]
\item   Any permutation $\sigma$ can be represented by an $N \times N$ {\em permutation matrix}
  $\matr{P}$, where $\matr{P}$ has  all entries equal to zero  except for $N$  entries equal to one
at the positions $(i,\sigma(i))$ for all $i\!\in \mathcal{N}$. 
\item Composing two permutations $\sigma$ and $\tau$ on $\mathcal{N}$ gives 
  two new permutations, $\sigma\tau$ and $\tau\sigma$, which in general are not equal. Equivalently, the product of two permutation matrices $\matr{P}$ and $\matr{Q}$ gives two new permutation matrices $\matr{PQ}$ and $\matr{QP}$, which in general are not equal. 
  
  \item (Permutation) matrices $\matr{P}$ and $\matr{Q}$ are said to have an \emph{overlapping column} (or row) if $\matr{P}$ and $\matr{Q}$ have at least one identical column (or row).  {Further, $\matr{P}$ and $\matr{Q}$ are said to be \emph{overlapping} if they have at least one overlapping column (or row), or \emph{non-overlapping} if they have no overlapping columns (or rows).}
  \item Matrix $\matr{P}$ is said to have a \emph{fixed column} (or row) if it overlaps with the identity matrix in at least one column (or row), or equivalently $\sigma$ has a \emph{fixed point} if $\sigma(i)=i$ for some  $i\!\in \mathcal{N}$.
  
  \item Two matrices $\matr{P}$ and $\matr{Q}$ \emph{commute} if $\matr{PQ}=\matr{QP}$. 
    \item Two (permutation) matrices $\matr{P}$ and $\matr{Q}$ are said to be \emph{strongly noncommutative} if $\matr{PQ}$ and $\matr{QP}$ have no overlapping columns, \emph{i.e.}, each column in $\matr{PQ}$ differs from the corresponding column in $\matr{QP}$. 

\end{itemize}
{\it {Circulant and circulant-block permutations}} 
\begin{itemize} 
\item The notation $\matr{I}_a^N$ is used to denote the $N\times N$ identity matrix with each row cyclically shifted to the left by $a$ positions. This matrix, and its corresponding permutation $\sigma$, will be referred to as a \emph{circulant} permutation matrix or permutation, respectively. 

\item  Let $a,b,m \in \mathbb{Z}$, $a,b \geq 0$, $m\geq 1$. Then 
\begin{propertyen} The circulant permutation matrix $\matr{I}_{a}^m$  has a fixed column iff $a \equiv 0\mod m$. {If $\mathbf{I}_a^m$ has a
fixed column, then $\mathbf{I}_a^m =\mathbf{I}_0^m$.}
\end{propertyen}
\begin{propertyen} The product of two circulant permutation matrices $\matr{I}_{a}^m$ and $\matr{I}_{b}^m$ is given by $\matr{I}_{a}^m\matr{I}_{b}^m=\matr{I}^m_{(a + b)\mod m}$.
\end{propertyen}\vspace{-5mm}
\begin{propertyen} The transposition of a circulant permutation matrix $\matr{I}_{a}^m$ is $\left(\matr{I}_{a}^m\right)^\tr=\matr{I}^m_{(m-a)\mod m}$.
\end{propertyen} 
\item By Property 2, any two circulant permutation matrices commute. This is not true for permutation matrices in general.

\item {We define an $mr\times mr$ \emph{circulant-block} permutation matrix $\matr{C}$ as an $m\times m$ array of $r \times r$ circulant permutation matrices and all-zero matrices arranged such that each (block) row and column contains precisely one circulant permutation matrix, \emph{i.e.},}
 \begin{equation}
        \matr{C} = \diag(\matr{I}^r_{s_{1}},\matr{I}^r_{s_{2}},\ldots,\matr{I}^r_{s_{m}})\cdot\tilde{\matr{P}},
       \end{equation}
{where $m$ and $r$ are positive integers,   $s_{k}\in[r]$, $k\in\{1,2,\ldots,m\}$, are called the \emph{shift parameters}, $\tilde{\matr{P}}\triangleq \matr{P}\otimes \matr{I}_0^r$ denotes the Kronecker product of an $m\times m$ permutation matrix $\matr{P}$ and $\matr{I}_{0}^r$, and, in a slight abuse of notation,}
\begin{align*}
& \diag(\matr{I}^r_{s_{1}},\matr{I}^r_{s_{2}},\ldots,\matr{I}^r_{s_{m}}) = \\&\left[\begin{array}{cccc}
 \matr{I}^r_{s_{1}} & \matr{0} & \cdots & \matr{0}\\
 \matr{0} &   \matr{I}^r_{s_{2}}& \cdots & \matr{0}\\
 \vdots &\vdots & \ddots & \vdots\\
  \matr{0} & \matr{0} & \cdots & \matr{I}^r_{s_{m}} 
\end{array}\right]_{mr \times mr}. 
\end{align*}
{The corresponding permutation $\sigma$ will be referred to as a \emph{circulant-block} permutation. }
\item {An example of an $mr\times mr$ circulant-block permutation matrix $\matr{C}$ with $m=4$, $r=7$, shift parameters \linebreak$(s_1,s_2,s_3,s_4)=(1,4,2,5)$, and $m \times m$ permutation matrix} $$\matr{P}=\left[\begin{array}{cccc}
 1 & 0 & 0 & 0\\
0 &   0& 0 & 1\\
0 &1 & 0& 0\\
 0 & 0& 1 &0
\end{array}\right]_{4\times 4},$$
{is} $$\matr{C}=\left[\begin{array}{cccc}
  \matr{I}^7_{1} & \matr{0} & \matr{0}  & \matr{0} \\
\matr{0}  &   \matr{0} & \matr{0}  & \matr{I}^7_{4}\\
\matr{0}  &\matr{I}^7_{2} & \matr{0} & \matr{0} \\
 \matr{0}  & \matr{0} & \matr{I}^7_{5} &\matr{0} 
\end{array}\right]_{21\times 21}.$$
\end{itemize} 

%%%%%%%%%%%%%%%%%%%%%%%%%%%%%%%%%%%%%%%%%%%%%%%%%%%

%\subsection{Protograph-based code construction}
%%%%%%%%%%%%%%%%%%%%%%%%%%%%%%%%%%%%%%%%%%%%%%%%%%%
\noindent {\it Protograph-based LDPC codes} 
\begin{itemize} 
\item A {\em protograph} \cite{tho03} is a small bipartite graph, represented by a parity-check or \emph{base} biadjacency matrix $\mathbf{B}$ (as  described in Section \ref{sec:intro}). 

\item The parity-check matrix $\mathbf{H}$ of a {\em protograph-based} LDPC block code is created by replacing each non-zero entry $ B_{i,j}$ in $\mathbf{B}$  by a {sum of $ B_{i,j}$ non-overlapping permutation matrices} of size $N\times N$ and each zero entry by the $N\times N$ all-zero matrix, where $B_{i,j}$ is a non-negative integer. 
\item  A parity-check matrix $\matr{H}$ that has been created from $\matr{B}$ using the protograph construction method with $N\times N$ permutation matrices is denoted by  $$\matr{H} = \matr{B}^{\uparrow N}.$$ 
\item Graphically, this operation is equivalent to taking an $N$-fold {\em graph cover} \cite{mas77}, or ``{\em $N$-lift}'', of the protograph. 

 \item An example of the lifting procedure applied to a $(3,4)$-regular protograph is shown in Fig. \ref{fig:tannergraph}. It is an important feature of this construction that each lifted code inherits the degree distribution and local graph neigbourhood structure of the protograph. 
\begin{figure}[h]
\begin{center}
\includegraphics[width=3in]{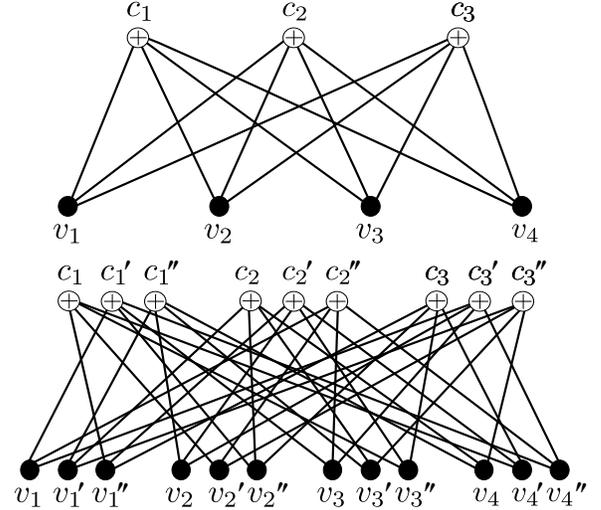}
\end{center}
\caption{Tanner graphs of a $(3,4)$-regular protograph (top) and a $(3,4)$-regular QC protograph-based code obtained from the protograph with $N=3$ (bottom).}\label{fig:tannergraph}
\end{figure} 
 \item The {\em ensemble of protograph-based LDPC codes} with block length $n = N n_v$, denoted $\xi_{\mathbf{B}}(N)$,  is defined as the set of matrices $\mathbf{H}$ that can be derived from a given base matrix $\base$ using all possible combinations of $N \times N$ permutation matrices.  
 \item The most general case of an LDPC code lifted from an $n_c\times n_v$ all-one base matrix is given by  a parity-check matrix $\matr{H} = \matr{B}^{\uparrow N}$ consisting of an $n_c\times n_v$ array of permutation matrices $\matr{Q}_{i,j}$, $i\in \{1,2,\ldots ,n_c\}$,  $ j\in \{1,2,\ldots ,n_v\}. $ Without loss of generality, after row and column permutations, any $n_c\times n_v$ all-one  base matrix  can be  written as
\begin{align} \label{matrix:general1}
&\begin{bmatrix} 
\matr{I}_0^N &\matr{I}_0^N &\ldots & \matr{I}_0^N \\
%I&P_{00} &\ldots& P_{0,n-2}\\
\matr{I}_0^N&\matr{P}_{2,2} &\ldots& \matr{P}_{2,n_v}\\ 
\vdots&\vdots&\ldots&\vdots\\
\matr{I}_0^N&\matr{P}_{n_c,2} &\ldots& \matr{P}_{n_c,n_v}
\end{bmatrix},
\end{align} 
where $\matr{P}_{i,j}$  is a permutation matrix, $i\in \{2,3,\ldots ,n_c\}$,  $ j\in \{2,3,\ldots ,n_v\}$,  $\matr{I}_0^N$ is the identity matrix, and all matrices are   of size $N$. The minimum distance and girth of the code and graph, respectively, are not affected by such operations. If all the permutation matrices $\matr{Q}_{i,j}$ are chosen to be circulant {(or circulant-block, as defined above)}, then the resulting permutation matrices $\matr{P}_{i,j}$ in \eqref{matrix:general1} are also circulant (resp. {circulant-block}). See Appendix \ref{sec:applift} for details.
%\item If all lifts are circulant, we obtain the {\em QC sub-ensemble}.... blah blah. 
\end{itemize}
{\it QC sub-ensembles} \begin{itemize} 
\item The {\em QC sub-ensemble}  of $\xi_{\mathbf{B}}(N)$, denoted $\xi^{QC}_{\mathbf{B}}(N)$, is the subset of parity-check matrices in $\xi_{\mathbf{B}}(N)$ where all the permutation matrices are chosen to be circulant. \item We denote a parity-check matrix $\matr{H}$ that has been $N$-lifted from $\matr{B}$ using \emph{only} circulant permutation matrices as  $$\matr{H} = \matr{B}^{\upmapsto N}.$$ 

\item The codes that are constructed using this technique are QC with \emph{period} $n_v$, \emph{i.e.}, cyclically shifting the $N$ symbols in each of the $n_v$ blocks in a codeword by one position {results in a codeword}.\footnote{Strictly speaking, for the code to be QC with period $n_v$, it must satisfy the property that, for each codeword, a cyclic shift of $n_v$ positions {results in a codeword}. This requires the columns of $\matr{H}$ to be reordered accordingly.}

\item {By restricting the choice of permutation matrices to come from the circulant subset $\{\matr{I}_a^N|a\in[N]\}$, the resulting  parity-check matrix $\mathbf{H}$ is the parity-check matrix of a QC-LDPC code}, \emph{i.e.}, $$\mathbf{H} = \matr{B}^{\upmapsto N}\in\xi^{QC}_{\mathbf{B}}(N)\subseteq\xi_{\mathbf{B}}(N).$$ In graphical terms, we refer to this operation as a ``circulant-based lifting''.
 \item Note that the sub-ensemble $\xi^{QC}_{\mathbf{B}}(N)$ is smaller than the ensemble $\xi_{\mathbf{B}}(N)$. This follows since there are only $N$ out of $N!$ permutations that are circulant, \emph{i.e.}, the fraction of choices of permutation matrices that are circulant is $N/N!=1/(N-1)!$, which tends to zero as $N\rightarrow \infty$. It follows that, if the base matrix $\mathbf{B}$ contains only ones and zeros, the fraction of codes in the ensemble that are composed of circulant matrices is $(1/(N-1)!)^t$, where $t$ is the number of ones in $\mathbf{B}$. Parallel edges in $\mathbf{B}$ further reduce this fraction. Consequently, asymptotic ensemble average results, such as those reported in \cite{ddja09,ady07,mpc13}, cannot be used to describe the behavior of this sub-ensemble, since the members are not \emph{typical}, \emph{i.e.}, the probability of picking such a code vanishes in the limit of large $N$, so we cannot say the codes perform close to the ensemble mean. 
\end{itemize} 
{\it QC-code examples}
\begin{example}\label{ex:simpleex}
{The $3\times 4$ all-ones base matrix} 
\begin{equation}\label{simple}\mathbf{B}=\left[\begin{array}{cccc}
                     1 & 1 & 1 & 1\\
                     1 & 1 & 1 & 1\\
                     1 & 1 & 1 & 1
                       \end{array}\right]\end{equation}  {can be lifted using circulant permutations with lifting factor $N=3$ to form the following $(3,4)$-regular QC-LDPC code with length $n=12$ and parity-check matrix}
                       \begin{equation} \mathbf{H}=  \matr{B}^{\upmapsto 3}=\left[\begin{array}{cccc}
                     \matr{I}_1^3 & \matr{I}_2^3 & \matr{I}_{1}^3 & \matr{I}_{2}^3 \\
                       \matr{I}_2^3 & \matr{I}_{1}^3 & \matr{I}_{2}^3 & \matr{I}_{0}^3 \\
 \matr{I}_{1}^3 & \matr{I}_{1} ^3 & \matr{I}_{1}^3 & \matr{I}_{2}^3\\
                       \end{array}\right]\in\xi^{QC}_{\mathbf{B}}(3).\end{equation}
The corresponding Tanner graphs are shown in Fig. \ref{fig:tannergraph}. \hfill$\Box$
\end{example}

\begin{example}\label{ex:tanex}
{The $(3,4)$-regular QC-LDPC Tanner code (see Example 11 in }\cite{sv12}{)} has a parity-check matrix, lifted from the $3\times 4$ all-ones base matrix $\base$, given by
\begin{equation}\label{tannercode}\mathbf{H} = \matr{B}^{\upmapsto N}=\left[\begin{array}{cccc}
                     \matr{I}_1^N & \matr{I}_2^N & \matr{I}_{4}^N & \matr{I}_{8}^N \\
                       \matr{I}_5^N & \matr{I}_{10}^N & \matr{I}_{20}^N & \matr{I}_{9}^N \\
 \matr{I}_{25}^N & \matr{I}_{19} ^N& \matr{I}_{7}^N & \matr{I}_{14}^N\\
                       \end{array}\right]\in\xi^{QC}_{\mathbf{B}}(N).\end{equation}
for the lifting factor $N=31$, this parity-check matrix defines a $[124,33,24]$ code with girth $8$.\hfill $\Box$
\end{example}
%When considering a sub-ensemble such as $\xi^{QC}_{\mathbf{B}}(N)$, one has to be careful with the relevance of asymptotic results obtained for the ensemble $\xi_{\mathbf{B}}(N)$. As $N\rightarrow \infty$, if the probablility of choosing a member of the sub-ensemble is non-zero we say that the code is a \emph{typical} member of the ensemble. By this definition, it is clear that the sub-ensemble $\xi^{QC}_{\mathbf{B}}(N)$ contains atypical codes. 
%\subsection
\\
{\it Minimum  distance bounds for QC sub-ensembles}
\begin{itemize}\item  If the base matrix $\mathbf{B}$ contains only ones and zeros, it is well known that the minimum distance of any code from the QC sub-ensemble of protograph-based LDPC codes can immediately be bounded above by $(n_c+1)!$ \cite{md01,fos04}.\end{itemize}
%\newtheorem{mddisttheorem}{\bf Theorem}
%\begin{mddisttheorem}[\hspace{0.01mm}{\cite[Theorem 2]{md01}}]
\begin{theorem}\label{thm:mddisttheorem}
 If a parity-check matrix of height $n_cM$ contains a submatrix of height $n_cM$ and width $(n_c+1)M$ {containing a grid of $n_c(n_c+1)$ permutation matrices} that all commute with each other, then the corresponding code has minimum distance less than or equal to $(n_c+1)!$. 
%\end{mddisttheorem}
\end{theorem}
\begin{itemize}\item In \cite{sv12}, the authors provide an improved bound that, in addition to giving tighter bounds for base matrices with only zero and one entries, can also be applied to base matrices with entries larger than one, \emph{i.e.}, protographs with parallel edges. Let the \emph{permanent} of an $m \times m$ matrix $\mathbf{M}$ be defined as
$$\mathrm{perm}(\mathbf{M}) = \sum_\sigma \prod_{x =1}^{m}M_{x,\sigma(x)},$$
where $M_{x,\sigma(x)}$ is the entry in $\matr{M}$ at position $(x,\sigma(x))$ and we sum over the $m!$ permutations $\sigma$ of the set $\{1,2,\ldots,m\}$. Then the minimum distance of a code drawn from the QC sub-ensemble of a protograph-based ensemble can be bounded above as follows.\end{itemize}
% \newtheorem{protobound}[mddisttheorem]{\bf Theorem}
%\begin{protobound}[\hspace{0.01mm}{\cite[Theorem 8]{sv12}}]
\begin{theorem}\label{thm:qcbound}
 Let $\cC$ be a code in $\xi^{QC}_{\mathbf{B}}(N)$, the QC sub-ensemble of the protograph-based ensemble of codes formed from a base matrix $\mathbf{B}$. Then the minimum Hamming distance of $C$ is bounded above as \begin{equation}\label{qcbound}
 \dmins \leq       \mathop{\mathrm{min}^*}_{\substack{S \subseteq \{1,2,\ldots,n_v\} \\ |S|=n_c+1}}    \sum_{i\in S}\mathrm{perm}(\mathbf{B}_{S\backslash i}),%\vspace{-1mm}
\end{equation}
where $\mathrm{perm}(\mathbf{B}_{S\backslash i})$ denotes the permanent of the matrix consisting of the $n_c$ columns of $\mathbf{B}$ in the set $S\backslash i$ and the $\mathrm{min}^*\{\cdot\}$ operator returns the smallest non-zero value from a set. 
%\end{protobound}  
\end{theorem}
\begin{itemize} \item For all the protographs considered in this paper, the bound on minimum distance obtained using (\ref{qcbound}) is at least as tight as $(n_c+1)!$, and in many cases it is tighter. Recently, Butler and Siegel further improved this bound for protographs with irregular structures and punctured symbols \cite{bs13}.\end{itemize}~\\
%\subsection
{\it Girth results for QC sub-ensembles}

 In this paper, our primary goal is to construct protograph-based QC-LDPC codes with large minimum distance; however, when using (sub-optimal) iterative decoding techniques, such as belief propagation (BP) decoding, graph-based properties, such as short cycles in the Tanner graph, are also important. Consequently, in order to achieve good decoding performance, we must ensure that we have an acceptable girth. Moreover, it is well known that there is a correspondence between short cycles in the Tanner graph and {low-weight codewords} for certain structured codes. In the following, we will consider $(2,K)$- and $(3,K)$-regular QC-LDPC codes. 
 \begin{itemize} \item For a $(2,K)$-regular code, $g=2\dmins$, since each codeword
corresponds directly to a cycle or a union of edge-disjoint
cycles \cite{wei04}. 
 \item {If $\matr{H}=\bcirc$ is lifted from the $3\times K$ all-ones base matrix $\base$, then we obtain the minimum distance bound  $\dmins\leq 24$ using Theorem }\ref{thm:mddisttheorem}{. In this case, the existence of a $4$- or $6$-cycle in the Tanner graph automatically implies a codeword of weight less than the upper bound $\dmins\leq 24$ (see Theorems 22 and 25 in }\cite{sv12}{). Consequently, a minimum girth of $8$ is required to achieve the minimum distance bound.}
\end{itemize} 
 It is well known that the girth of the Tanner graph associated with a parity-check matrix composed of circulant permutation matrices can be determined quickly using modular arithmetic \cite{fos04}, \cite{osu06}. In \cite{smc11}, a technique was presented to derive a set of conditions on the permutation matrices of a protograph-based parity-check matrix $\matr{H}=\matr{B}^{\uparrow N}\in\xi_{\mathbf{B}}(N)$ in order to achieve a certain desired girth $g$. It was shown that, if certain products of the permutation matrices comprising $\matr{H}$ do not have any fixed columns, then the girth will be at least $g$. In this paper, we construct protograph-based QC-LDPC codes with large minimum distance and use these conditions on the permutation matrices in order to achieve a certain guaranteed girth. 

%%%%%%%%%%%%%%%%%%%%%%%%%%%%%%%%%%%%%%%%%%%%%%%%%%%
%%%%%%%%%%%%%%%%%%%%%%%%%%%%%%%%%%%%%%%%%%%%%%%%%%%
\section{Protograph-based LDPC codes obtained by ``pre-lifting" a protograph: design and analysis}\label{sec:prelifting}
%%%%%%%%%%%%%%%%%%%%%%%%%%%%%%%%%%%%%%%%%%%%%%%%%%%
%%%%%%%%%%%%%%%%%%%%%%%%%%%%%%%%%%%%%%%%%%%%%%%%%%%
In this section, we introduce a  two-step lifting procedure based on a protograph.  %a ``pre-lifting'' step where we take an $m$-fold graph cover of the protograph, where $m$ is typically small, and a ``second-lifting'' step where we take an $r$-fold graph cover of the \emph{pre-lifted} protograph, where $r$ is typically large and the permutations are chosen to be cyclic. 
Based on this procedure, we describe how to  construct QC-LDPC codes with increased girth and minimum distance while maintaining the circulant-based structure that facilitates efficient implementation. 
%In particular, we show that the QC-LDPC ensemble obtained from a pre-lifted protograph can have an increased upper bound on minimum distance compared to the QC-LDPC ensemble obtained from the original protograph and we demonstrate the existence of codes with minimum distance exceeding the original bound. We also present two design rules for code construction: one uses only commuting pairs of permutation matrices at the first (pre-lifting) stage, while the other uses at least one pair of non-commuting permutation matrices. In each case, we obtain a significant increase in the minimum distance compared to a one-step circulant-based lifting of the original protograph and achieve a certain guaranteed girth. The expected performance improvement is verified by simulation results.
In the following, we mostly focus on base matrices $\matr{B}$ with only zero and one entries, \emph{i.e.}, protographs without parallel edges. This assumption simplifies analysis and ensures that the resulting codes are amenable to low-complexity implementation. (We demonstrate in Section \ref{sec:multi} that the technique can also be successfully applied to base matrices with parallel edges.) 
%\begin{itemize}
%\item If entry $B_{i,j}$ of $\matr{B}$ is equal to one, then the corresponding block of the lifted parity-check matrix consists of an $N\times N$ permutation matrix $\matr{P}_{i,j}$.  
%\item The minimum distance of any code $C$ derived from $\matr{B}$, where the permutation matrices $\matr{P}_{i,j}$ are chosen to be circulant, is bounded above by $(n_c+1)!$ for an \emph{arbitrarily large} lifting factor $N$ (see Theorem \ref{thm:mddisttheorem}). 
%\end{itemize}
%We will show that by choosing the permutation matrices $\matr{P}_{i,j}$ to be composed of a sub-array of $r\times r$ smaller circulant matrices, we can derive QC codes with minimum distance exceeding this upper bound. 
%%%%%%%%%%%%%%%%%%%%%%%%%%%%%%%%%%%%%%%%%%%%%%%%%%%
\subsection{{Constructing QC-LDPC codes by prelifting}}

The construction technique can be defined in two steps:
\begin{enumerate}
 \item first, a ``{\bf  pre-lifting}'' step where we take a carefully chosen $m$-fold graph cover of the protograph with base matrix $\matr{B}=[B_{i,j}]_{n_c\times n_v}$, where $m$ is typically small, to form a {\em pre-lifted base matrix} $$\matr{B}^{\uparrow m}=[\matr{B}_{i,j}],$$ where $\matr{B}_{i,j}$ is an $m\times m$ permutation matrix if $B_{i,j}=1$, or the $m\times m$ all zero matrix if $B_{i,j}=0$,
\item following this,  a second {\bf $r$-fold lifting} step where we take an $r$-{\em fold  graph cover} of the \emph{pre-lifted protograph} associated with $\matr{B}^{\uparrow m}$, where $r$ is typically large. The permutations are chosen to be circulant, creating a QC-LDPC code with parity-check matrix \begin{equation}\matr{H}=\matr{B}^{\uparrow m\upmapsto r}=[\matr{H}_{i,j}],\end{equation}
where  \begin{equation}\label{Pij}\matr{H}_{i,j}=(\matr{B}_{i,j})^{\upmapsto r}\end{equation} is an $mr\times mr$ {circulant-block permutation matrix (see Section} \ref{sec:background}). %, which can be described as the product
% \begin{equation}\label{Pij}
  %      \matr{H}_{i,j} = \diag(\matr{I}^r_{s_{i,j,1}},\matr{I}^r_{s_{i,j,2}},\ldots,\matr{I}^r_{s_{i,j,m}})\cdot\tilde{\matr{B}}_{i,j},
  %     \end{equation}
%where the \emph{shift parameters} $s_{i,j,k}\in[r]$, $k\in\{1,2,\ldots,m\}$,  $\tilde{\matr{B}}_{i,j}\triangleq \matr{B}_{i,j}\otimes \matr{I}_0^r$ denotes the Kronecker product of matrices $\matr{B}_{i,j}$ and $\matr{I}_{0}^r$, and, in a slight abuse of notation, 
%\begin{align*}
%& \diag(\matr{I}^r_{s_{i,j,1}},\matr{I}^r_{s_{i,j,2}},\ldots,\matr{I}^r_{s_{i,j,m}}) = \\&\left[\begin{array}{cccc}
 %\matr{I}^r_{s_{i,j,1}} & \matr{0} & \cdots & \matr{0}\\
% \matr{0} &   \matr{I}^r_{s_{i,j,2}}& \cdots & \matr{0}\\
% \vdots &\vdots & \ddots & \vdots\\
%  \matr{0} & \matr{0} & \cdots & \matr{I}^r_{s_{i,j,m}} 
%\end{array}\right].
%\end{align*}

\end{enumerate}
The codes that are constructed using this technique are QC with period $mn_v$, \emph{i.e.}, cyclically shifting the $r$ symbols in each of the $mn_v$ blocks in a codeword by one position results in another codeword.
\subsection{Examples of pre-lifting}
\begin{example}\label{ex:23prelift}
%To demonstrate the procedure, 
Consider the $(2,3)$-regular base matrix 
\begin{equation*}
 \matr{B}=\left[\begin{array}{ccc}
 1 & 1 & 1\\
 1 & 1 & 1
\end{array}\right].
\end{equation*}
%We find that 
\begin{itemize} \item ({\em One-step circulant lifting}) Any QC-LDPC code derived from $\matr{B}$ using a one-step circulant-based lifting, \emph{i.e.}, with parity-check matrix $$\matr{H}=\matr{B}^{\upmapsto N}=\left[\begin{array}{ccc}
 \matr{I}_{a}^N & \matr{I}_{b}^N & \matr{I}_{c}^N\\
 \matr{I}_{d}^N & \matr{I}_{e}^N & \matr{I}_{f}^N
\end{array}\right]\in\xi_{\matr{B}}^{QC}(N)\subseteq \xi_{\mathbf{B}}(N),$$ has its minimum distance upper bounded by $(n_c+1)!=6$ and its girth upper bounded by $12$. (Recall that, for a parity-check matrix with column weight $2$, $g=2\dmins$, since each codeword
corresponds directly to a cycle or a union of edge-disjoint
cycles.) 
\item ({\em Pre-lifting}) A pre-lifted QC-LDPC code is obtained from $\matr{B}$ using \begin{itemize}  \item a pre-lifted base matrix of the form
\begin{equation*}
\matr{B}^{\uparrow m}=\left[\begin{array}{ccc}
 \matr{B}_{1,1} & \matr{B}_{1,2} & \matr{B}_{1,3}\\
 \matr{B}_{2,1} & \matr{B}_{2,2} & \matr{B}_{2,3}
\end{array}\right]\in\xi_{\mathbf{B}}(m),
\end{equation*}
where each $\matr{B}_{i,j}$ is an $m \times m$ permutation matrix. \item an $r$-lifting of $\matr{B}^{\uparrow m}$ to  $\matr{B}^{\uparrow m\upmapsto r}$ to obtain\begin{align*}
 \matr{H}&=\matr{B}^{\uparrow m\upmapsto r}=%\\&
 \left[\begin{array}{ccc}
 \matr{H}_{1,1} & \matr{H}_{1,2} & \matr{H}_{1,3}\\
 \matr{H}_{2,1} & \matr{H}_{2,2} & \matr{H}_{2,3}
\end{array}\right]\\&\in\xi_{\bpre}^{QC}(r) \subseteq \xi_{\mathbf{B}}(mr),
\end{align*}
where each $\matr{H}_{i,j}$ is obtained by replacing every one in $\matr{B}_{i,j}$ with an $r \times r$ circulant permutation matrix.
\end{itemize}
\item ({\em Numerical pre-lifting example}) %For example,
 Consider the following pre-lifted base matrix with $m=2$: \begin{align}\label{prelift23}\nonumber
 \matr{B}^{\uparrow 2}= &\left[\begin{array}{ccc}
 \matr{B}_{1,1} & \matr{B}_{1,2} & \matr{B}_{1,3}\\
 \matr{B}_{2,1} & \matr{B}_{2,2} & \matr{B}_{2,3}
\end{array}\right] \\=&
\left[\begin{array}{cc|cc|cc}
1 & 0 & 1 & 0 & 1 & 0\\
0 & 1 & 0 & 1 & 0 & 1\\\hline
1 & 0 & 1 & 0 & 0 & 1\\
0 & 1 & 0 & 1 & 1 & 0\end{array}\right] \in\xi_{\mathbf{B}}(2).
\end{align}
 Any code drawn from the QC-LDPC code ensemble based on this pre-lifted base matrix $\matr{B}^{\uparrow 2}$ has its minimum distance and girth bounded above by $10$ and $20$, respectively, which exceeds the upper bounds associated with the original base matrix $\matr{B}$. The following circulant-based lifting of $\matr{B}^{\uparrow 2}$ with $r=20$, 
\begin{align*}
 \matr{H}&=\matr{B}^{\uparrow 2 \upmapsto 20} =\left[\begin{array}{ccc}
 \matr{H}_{1,1} & \matr{H}_{1,2}& \matr{H}_{1,3}\\
 \matr{H}_{2,1} & \matr{H}_{2,2} & \matr{H}_{2,3}
\end{array}\right] \\&=
\left[\begin{array}{cc|cc|cc}
\matr{I}_0^{20} & \matr{0} & \matr{I}_0^{20} & \matr{0} & \matr{I}_0^{20} & \matr{0}\\
\matr{0} & \matr{I}_0^{20} & \matr{0} & \matr{I}_0^{20} & \matr{0} & \matr{I}_0^{20}\\\hline
\matr{I}_0^{20} & \matr{0} & \matr{I}_1^{20} & \matr{0} & \matr{0} & \matr{I}_0^{20}\\
\matr{0} & \matr{I}_0^{20} & \matr{0} & \matr{I}_9^{20}& \matr{I}_4^{20} & \matr{0}\end{array}\right]\\&\in \xi_{\mathbf{B}^{\uparrow 2}}^{QC}(20)\subseteq \xi_{\mathbf{B}}(40),
\end{align*}
%$\matr{H}\in \xi_{\mathbf{B^\prime}}^{QC}(20)$, 
defines a $[120,41,10]$ QC code with girth $g =20$, \emph{i.e.}, it achieves the improved upper bounds.\footnote{The parity-check matrix $\matr{H}$ has rank $79$, and hence the dimension of the code is $k=41$.} 
Note that 
$$ \matr{H}=\left[\begin{array}{ccc}
 \matr{I}_{0}^{40} & \matr{I}_{0}^{40} & \matr{I}_{0}^{40}\\
 \matr{I}_{0}^{40} & \matr{P} & \matr{Q}
\end{array}\right],$$
but the permutation matrices $\matr{P}$ and $\matr{Q}$ are not circulant, \emph{i.e.}, they cannot be written in the form $ \matr{I}_{a}^{40}$ for some integer $a$. Thus the bound $\dmins \leq 6$ does not apply.\hfill $\Box$\end{itemize}
\end{example}

%\newtheorem{preliftremark}[mddisttheorem]{\bf Remark}
%\begin{preliftremark}
\begin{remark}\label{thm:preliftremark}
{\em Clearly, the pre-lifted base matrix $\matr{B}^{\uparrow m}$ defines a code that exists in the ensemble of all codes lifted from $\matr{B}$ with lifting factor $m$,  $\xi_{\mathbf{B}}(m)$, and the QC code with parity-check matrix $\matr{H}=\matr{B}^{\uparrow m\upmapsto r}$ obtained after the circulant lifting step exists in $\xi_{\mathbf{B}}(mr)$; however, $\matr{H}$ does not necessarily exist in $\xi^{QC}_{\mathbf{B}}(mr)$, and thus the minimum distance may exceed $(n_c+1)!$. Note that, since $\matr{H}\in\xi_{\mathbf{B}}(mr)$, the resulting code preserves the local graph neigbourhood structure and degree distribution of the protograph. Moreover, because $\matr{H}$ is composed of circulants, we maintain the efficient implementation advantages of QC codes.}
%\end{preliftremark}
\end{remark}
Our goal in this paper is to study this two-step lifting process and determine how to construct QC-LDPC codes based on protographs with improved minimum distance and a certain guaranteed girth compared to one-step circulant-based liftings. In the following, we investigate the effect of pre-lifting a protograph on girth and minimum distance.
%%%%%%%%%%%%%%%%%%%%%%%%%%%%%%%%%%%%%%%%%%%%%%%%%%%
%\subsection{Properties of pre-lifted photographs} 

\subsection{Girth properties  of pre-lifted protographs}\label{sec:preliftgirth}
%%%%%%%%%%%%%%%%%%%%%%%%%%%%%%%%%%%%%%%%%%%%%%%%%%%

In this section, we will establish some results on the girth of a parity-check matrix obtained from a pre-lifted base matrix $\bpre$. These results will later be used to obtain pre-lifted QC-LDPC codes with a certain desired girth. This is important because short cycles have an adverse effect when decoding LDPC codes using iterative BP decoding.  Also, there is a close connection between short cycles and {low-weight codewords}. In this regard, the structure imposed by a protograph is important. 

It is well known that any cycle in a graph cover can be mapped to a cycle in the base graph (or protograph). As a direct consequence, we state the following result.
\begin{lemma} \label{thm:girthremark}
{If a protograph has girth $g$, then the girth of any $N$-lifted graph is bounded below by $g$.}
\end{lemma}~ \\

Theorem \ref{thm:girthremark} implies the following corollary concerning the girth of a pre-lifted base matrix.

\begin{corollary} \label{thm:girthcorollary}
If a pre-lifted base matrix $\bpre$ has girth $g$, then the girth of any code from the ensemble $\xi_{\bpre}^{QC}(r)$ is bounded below by $g$, for any lifting factor $r$.
\end{corollary}~ \\

 {\em Design implications of Corollary \ref{thm:girthcorollary}:} 
 One could use a technique such as progressive edge growth (PEG) \cite{hea05} to design a pre-lifted base graph with girth $g$ and then, by using circulant permutations at the second lifting step, construct a QC-LDPC code with girth at least as large as $g$. However, even if the permutation matrices $\mathbf{B}_{i,j}$ chosen in the first lifting step do not  satisfy the conditions needed to  guarantee girth $g$, these conditions can still be satisfied in the second lifting step by {carefully choosing the circulant matrices $\matr{I}^r_{s_{i,j,k}}$ comprising $\matr{H}_{i,j}$}. Moreover, obtaining girth $g$ in the first lifting step will typically require a large lifting factor $m$, where we want $m$ to be as small as possible in order to simplify the analysis and implementation. An example is given in Section \ref{sec:circperm}.  

If we wish to increase the girth from that of the pre-lifted base matrix $\bpre$, it is necessary to check if certain products of the {circulant-block} permutation matrices comprising $ \matr{H}=\matr{B}^{\uparrow m \upmapsto r}$ have fixed columns (see Section \ref{sec:3x4}). The following lemma proves useful to reduce the number of such conditions that one needs to check.

%\newtheorem{girthlemma}[mddisttheorem]{\bf Lemma}
%\begin{girthlemma}
\begin{lemma}\label{thm:girthlemma}
Let $\matr{P}$ and $\matr{Q}$ be two $mr\times mr$ {circulant-block} permutation matrices derived from $m\times m$ permutation matrices $\matr{B}_{P}$ and $\matr{B}_{Q}$, respectively. Then the product $\matr{P}\matr{Q}$ cannot have a fixed column if $\matr{B}_{P}\matr{B}_{Q}$ does not have a fixed column.
%\end{girthlemma}
\end{lemma}
\emph{Proof}. See Appendix \ref{sec:appa}. 

\subsection{Minimum distance properties of pre-lifted protograph-based codes}\label{sec:MacKay}
%%%%%%%%%%%%%%%%%%%%%%%%%%%%%%%%%%%%%%%%%%%%%%%%%%%
In \cite{md01}, MacKay and Davey established that, for a parity-check matrix with {an $n_c\times n_v$ grid of commuting permutation matrices}, the minimum distance is bounded above by $(n_c+1)!$ (cf. Theorem \ref{thm:mddisttheorem}). We now establish a similar result for a grid of  commuting {circulant-block} permutation matrices based on a pre-lifted base matrix $\bpre$. To prove this, we require the following Lemma.
%\newtheorem{disttheorem}[mddisttheorem]{\bf Theorem}
%\begin{disttheorem}
\begin{lemma}\label{thm:distlemma}
Suppose that two {circulant-block} permutation matrices are given as
\begin{eqnarray*}
 \matr{P}&=&\diag(\matr{I}^r_{p_1},\matr{I}^r_{p_1},\ldots,\matr{I}^r_{p_1})\cdot\tilde{\matr{B}}_P=\matr{B}_P\otimes\matr{I}^r_{p_1},\\
\matr{Q}&=&\diag(\matr{I}^r_{q_1},\matr{I}^r_{q_1},\ldots,\matr{I}^r_{q_1})\cdot\tilde{\matr{B}}_Q=\matr{B}_Q\otimes\matr{I}^r_{q_1},
\end{eqnarray*}
where $p_1,q_1\in[r]$  and $\matr{B}_P$ and $\matr{B}_Q$ are $m\times m$ permutation matrices. Then, $$\matr{PQ}=\matr{QP}\quad \quad   {\rm iff} \quad \quad \matr{B}_P\matr{B}_Q=\matr{B}_Q\matr{B}_P.$$
%\end{distlemma}
\end{lemma}
\emph{Proof}. See Appendix \ref{sec:appb}. 

Then, the main result follows. 
\begin{theorem}\label{thm:disttheorem}
Let $\bpre$ be a pre-lifted base matrix derived from an $n_c\times n_v$ binary base matrix $\matr{B}$, and suppose  $$\matr{B}_{i,j}\matr{B}_{k,l}=\matr{B}_{k,l}\matr{B}_{i,j},$$ for all $i,k\in \{1,2,\ldots,n_c\}, j,l\in \{1,2,\ldots,n_v\}$, $(i,j)\neq (k,l)$. If $$s_{i,j,1}=s_{i,j,2}=\cdots=s_{i,j,m},$$ for each {circulant-block} permutation matrix $\matr{H}_{i,j}$, {as defined in (}\ref{Pij}{)}, then the minimum distance of any code $C\in\xi_{\bpre}^{QC}(r)$ is bounded above by $(n_c+1)!$.%, \emph{i.e.}, the upper bound is equal to that associated with the original base matrix $\matr{B}$.
%\end{disttheorem} 
\end{theorem}
\emph{Proof}. By applying Lemma \ref{thm:distlemma} to each pair of {circulant-block} permutation matrices $(\matr{H}_{i,j},\matr{H}_{k,l}),$ $\forall i,k\in \{1,2,\ldots,n_c\}$, $j,l\in \{1,2,\ldots,n_v\}$, corresponding to the pair $(\matr{B}_{i,j}\matr{B}_{k,l})$, we find that all pairs of matrices commute and thus the {result of Theorem }\ref{thm:mddisttheorem}{ holds.}\hfill $\Box$~\\\\
{\em Design Implications of Theorem \ref{thm:disttheorem}:} 
In order to have minimum distance exceeding $(n_c+1)!$, we must have at least one pair of non-commuting {circulant-block} permutation matrices. In fact, we require that at least one pair of {circulant-block} matrices {is strongly noncommutative}. Note that, in general, if%\vspace{-2mm}
\begin{eqnarray*}
 \matr{P}&=&\diag(\matr{I}_{p_1}^r,\matr{I}_{p_2}^r,\ldots,\matr{I}_{p_m}^r)\cdot\tilde{\matr{B}}_P,\label{P}\\
\matr{Q}&=&\diag(\matr{I}_{q_1}^r,\matr{I}_{q_2}^r,\ldots,\matr{I}_{q_m}^r)\cdot\tilde{\matr{B}}_Q\label{Q},
\end{eqnarray*}%\vspace{-3mm}
then, as described in the proof of Lemma \ref{thm:girthlemma}, %\vspace{-1mm}
%\begin{equation*}
%\begin{split}
% \matr{PQ}=&\diag(\matr{I}_{p_1},\matr{I}_{p_2},\ldots,\matr{I}_{p_m})\cdot\tilde{\matr{B}}_P\cdot\\&\hspace{1mm}\diag(\matr{I}_{q_1},\matr{I}_{q_2},\ldots,\matr{I}_{q_m})\cdot\tilde{\matr{B}}_Q
%\end{split}
% \matr{PQ}=\diag(\matr{I}_{p_1}^r,\matr{I}_{p_2}^r,\ldots,\matr{I}_{p_m}^r)\cdot\tilde{\matr{B}}_P\cdot\diag(\matr{I}_{q_1}^r,\matr{I}_{q_2}^r,\ldots,\matr{I}_{q_m}^r)\cdot\tilde{\matr{B}}_Q%\vspace{-1mm}
%\end{equation*}
%and %\vspace{-1mm}
%\begin{equation*}
%\begin{split}
%\tilde{\matr{B}}&_P\cdot\diag(\matr{I}_{q_1},\matr{I}_{q_2},\ldots,\matr{I}_{q_m})\\&= \diag(\matr{I}_{q_{\sigma(1)}},\matr{I}_{q_{\sigma(2)}},\ldots,\matr{I}_{q_{\sigma(m)}})\cdot\tilde{\matr{B}}_P,
%\end{split}
%\tilde{\matr{B}}_P\cdot\diag(\matr{I}_{q_1}^r,\matr{I}_{q_2}^r,\ldots,\matr{I}_{q_m}^r)= \diag(\matr{I}^r_{q_{\sigma(1)}},\matr{I}^r_{q_{\sigma(2)}},\ldots,\matr{I}^r_{q_{\sigma(m)}})\cdot\tilde{\matr{B}}_P,%\vspace{-1mm}
%\end{equation*}
%where $\sigma$ is the permutation associated with $\matr{B}_P$. Consequently,%\vspace{-1mm}
{ \begin{align}
 \matr{PQ}=\diag(\matr{I}^r_{p_1+q_{\sigma(1)}},\matr{I}^r_{p_2+q_{\sigma(2)}},\ldots,\matr{I}^r_{p_m+q_{\sigma(m)}})\cdot\tilde{\matr{B}}_P\tilde{\matr{B}}_Q,\label{commute1}\\
\matr{QP}=\diag(\matr{I}^r_{q_1+p_{\tau(1)}},\matr{I}^r_{q_2+p_{\tau(2)}},\ldots,\matr{I}^r_{q_m+p_{\tau(m)}})\cdot\tilde{\matr{B}}_Q\tilde{\matr{B}}_P,\label{commute2}
\end{align}}%
%$$\begin{array}{c}
% \matr{PQ}=\diag(\matr{I}_{p_1+q_{\sigma(1)}},\matr{I}_{p_2+q_{\sigma(2)}},\ldots,\matr{I}_{p_m+q_{\sigma(m)}})\cdot\tilde{\matr{B}}_P\cdot\tilde{\matr{B}}_Q,\\
%\matr{QP}=\diag(\matr{I}_{q_1+p_{\tau(1)}},\matr{I}_{q_2+p_{\tau(2)}},\ldots,\matr{I}_{q_m+p_{\tau(m)}})\cdot\tilde{\matr{B}}_Q\cdot\tilde{\matr{B}}_P,\end{array}\vspace{-1mm}$$
\noindent where $\sigma$ and $\tau$ are the permutations associated with $\matr{B}_P$ and $\matr{B}_Q$, respectively, and addition is performed modulo $r$. In addition, %\vspace{-1mm}
\begin{align*}
\tilde{\matr{B}}_P\tilde{\matr{B}}_Q =\left(\matr{B}_P\otimes \matr{I}_0^r\right)\left(\matr{B}_Q\otimes \matr{I}^r_0\right)=\matr{B}_P\matr{B}_Q\otimes \matr{I}_0^r\matr{I}_0^r  %\vspace{-2mm}  
                                                                                                                                                                                                                                 \end{align*}
by the distributive law of the Kronecker product, and it follows that \begin{equation}\label{commuteiff}
\tilde{\matr{B}}_P\tilde{\matr{B}}_Q=\tilde{\matr{B}}_Q\tilde{\matr{B}}_P \iff \matr{B}_P\matr{B}_Q=\matr{B}_Q\matr{B}_P.\end{equation}~\\
Consequently, to ensure $\matr{P}$ and $\matr{Q}$ are strongly noncommutative, we consider two cases:
\begin{itemize}
\item if $\matr{B}_P\matr{B}_Q=\matr{B}_Q\matr{B}_P$, then we see from (\ref{commute1}), (\ref{commute2}), and (\ref{commuteiff}) that we must 
ensure that the diagonal matrices in (\ref{commute1}) and (\ref{commute2}) do not have an overlapping column; 
\item if $\matr{B}_P$ and $\matr{B}_Q$ are strongly noncommutative, then $\matr{P}$ and $\matr{Q}$ are also strongly noncommutative, even if the diagonal matrices in  (\ref{commute1}) and (\ref{commute2}) are equal. 
\end{itemize}
In the next section, we will use these two cases to propose two new design rules for constructing QC-LDPC codes based on a pre-lifted protograph.
\subsection{Designing good pre-lifted protographs}
%%%%%%%%%%%%%%%%%%%%%%%%%%%%%%%%%%%%%%%%%%%%%%%%%%%
%\subsection
% for pre-lifted protograph-based QC-LDPC codes}\label{sec:design}
%%%%%%%%%%%%%%%%%%%%%%%%%%%%%%%%%%%%%%%%%%%%%%%%%%%

In order to avoid being constrained by the upper bound of Theorem \ref{thm:disttheorem}, it is necessary to ensure that there is at least one pair of strongly noncommutative {circulant-block} permutation matrices in $\matr{H}$ (see the discussion of Theorem \ref{thm:disttheorem} in Section~\ref{sec:MacKay}). We now provide two new design rules for constructing QC-LDPC codes based on a pre-lifted protograph depending on whether the permutation matrices used for pre-lifting commute or not.
\begin{itemize}
 \item  
{\bf {Design Rule 1}:}  {\em Commuting pre-lifting permutation matrices.} In this case, at Step 1, each pair of matrices $\matr{B}_{i,j}$ and $\matr{B}_{k,l}$, $(i,j)\neq(k,l)$, is chosen to be commuting. (Typically, we choose circulant matrices in applying {Design Rule 1}, since they necessarily commute.)  At the second step, since the pre-lifting permutation matrices commute, we must ensure that the diagonal matrices are chosen such that at least one pair of {circulant-block} matrices $(\matr{P},\matr{Q})$ in $\matr{H}$ {is strongly noncommutative}, \emph{i.e.},  $$p_i+q_{\sigma(i)}\not\equiv q_i+p_{\tau(i)}\mod r,$$ for all $i\in \{1,2,\ldots,m\}$ (thus necessarily $\matr{PQ}\neq\matr{QP}$). This can be achieved, for example,  by imposing the condition that $\tau$ has no fixed point, setting $q_1=q_2=\cdots=q_m$, and choosing each $p_i$ to be distinct.
\item 

{\bf {Design Rule 2}:} {\em  Non-commuting pre-lifting permutation matrices.}
In Step 1, we choose permutation matrices $\matr{B}_{i,j}$ and ensure that at least one pair of matrices $(\matr{B}_{i,j},\matr{B}_{k,l})$, $(i,j)\neq(k,l)$, {is strongly noncommutative}, thus necessarily $$\matr{B}_{i,j}\matr{B}_{k,l}\neq\matr{B}_{k,l}\matr{B}_{i,j}.$$ {At Step 2, we then choose all circulant permutation matrices in each circulant-block matrix to have the same shift parameter}, \emph{i.e.}, $p_1=p_2=\cdots=p_m$ for {circulant-block} $\matr{P}$.
\end{itemize}
%\subsection{Remarks on the design rules} 
%\newtheorem{designremark}[mddisttheorem]{\bf Remark}
%\begin{designremark}
%$\begin{remark}\label{thm:designremark}
%{\em 
These design rules give necessary (but not sufficient) conditions for pre-lifted QC-LDPC codes to have minimum distance exceeding that of QC-LDPC codes lifted directly from $\matr{B}$. Note that the rules above apply directly when $n
_v=n_c+1$; however, when $n_v>n_c+1$, they must be applied to every $n_c\times (n_c+1)$ block submatrix.

We will later give examples of how the permutations at both steps should be chosen to ensure large minimum distance and girth. In Section \ref{sec:constructingprelifted}, we focus on pre-lifting (Step 1) and discuss choosing permutations to maximize the distance upper bound calculated from (\ref{qcbound}). At the pre-lifting step, the conditions on the circulants that must be checked to guarantee a desired girth $g$ at the next step can be determined, and we demonstrate in Section \ref{sec:liftcirculant} that certain choices of pre-lifting can reduce the number of conditions to be checked or even eliminate the need to check any conditions. For both design rules, we then provide examples in Sections \ref{sec:constructingprelifted}-\ref{sec:liftnoncommuting} of circulants chosen at Step 2 that result in improved minimum distance and achieve a desired girth $g$.%}%\end{designremark}
%\end{remark} 
%%%%%%%%%%%%%%%%%%%%%%%%%%%%%%%%%%%%%%%%%%%%%%%%%%%

%\newtheorem{distremark}[mddisttheorem]{\bf Remark}
%\begin{distremark}
\begin{remark}\label{thm:distremark}
{\em  {By ensuring that some pairs of circulant-block permutation matrices $\matr{H}_{i,j}$ are strongly noncommutative}, we can construct pre-lifted QC-LDPC codes with minimum distance exceeding $(n_c+1)!$. This can be observed by examining the proof of Theorem 2 in \cite{md01}. When some pairs of permutation matrices are strongly noncommutative, instead of finding a codeword of weight $(n_c+1)!$, we obtain a binary vector of weight $(n_c+1)!$ that has a small number  $f>0$ of unsatisfied parity-check equations (commonly referred to as a $((n_c+1)!,f)$ \emph{near-codeword}). An example is given in  Appendix \ref{sec:appc}.}
%\end{distremark}
\end{remark}
%\newtheorem{searchremark}[mddisttheorem]{\bf Remark}
%\begin{searchremark}
\begin{remark}\label{thm:searchremark}
{\em In general, when constructing short to moderate length parity-check matrices $\matr{H} =  [\matr{H}_{i,j}]$, it is a difficult problem to search for permutation matrices $\matr{H}_{i,j}$ such that the code achieves a desired minimum distance and girth. We will see in the following sections that this search is much simpler if we construct the parity-check matrices using a two-step method and {circulant-block} permutation matrices $\matr{H}_{i,j}$}.\end{remark}
%\end{searchremark}
%%%%%%%%%%%%%%%%%%%%%%%%%%%%%%%%%%%%%%%%%%%%%%%%%%%
\section{Code design: Selecting \\permutations for pre-lifting}\label{sec:constructingprelifted}
%%%%%%%%%%%%%%%%%%%%%%%%%%%%%%%%%%%%%%%%%%%%%%%%%%%
%%%%%%%%%%%%%%%%%%%%%%%%%%%%%%%%%%%%%%%%%%%%%%%%%%%

In this section, we focus on the selection process for the permutations involved in the first step of the construction technique by considering two examples: a simple $(2,3)$-regular protograph that is useful in describing the method and is easy to analyze, and a more practically interesting $(3,4)$-regular protograph that demonstrates the successful application of the method to a protograph with larger node degrees. 

\begin{itemize}
\item  For the first example, we show that the upper bounds on minimum distance and girth obtained for the original base matrix can be increased by pre-lifting the protograph and that the new upper bounds increase as larger degrees of pre-lifting are considered. We demonstrate that the improved minimum distance and girth promised by the increased upper bounds are indeed obtainable by selecting appropriate circulants at the second lifting step and we give explicit constructions showing the increased bounds are in fact tight.

\item For the second example, we show that even larger gains in minimum distance are possible. In particular, we show that the upper bound on minimum distance can be increased significantly by pre-lifting and confirm the improvement by providing specific constructions with improved minimum distance (larger than the original upper bounds) and a certain guaranteed girth. 
\end{itemize}
%%%%%%%%%%%%%%%%%%%%%%%%%%%%%%%%%%%%%%%%%%%%%%%%%%%
\subsection{Pre-lifted QC structures for a $2 \times 3$ base matrix}\label{sec:2x3}
%%%%%%%%%%%%%%%%%%%%%%%%%%%%%%%%%%%%%%%%%%%%%%%%%%%

We begin our study with a base matrix of column weight $2$; in particular, the $(2,3)$-regular  base matrix $\matr{B}$ discussed in Example \ref{ex:23prelift}. 
\begin{itemize}\item Any $N$-fold graph cover of $\matr{B}$ can be written in the form of \eqref{matrix:general1}, \emph{i.e.},
\begin{align} \matr{H}=\matr{B}^{\uparrow N}=\label{matrix23gen}
&\begin{bmatrix} \matr{I}_0^N &\matr{I}_0^N &\matr{I}_0^N\\\matr{I}_0^N&\matr{P} & \matr{Q}
\end{bmatrix}\in \xi_{\matr{B}}(N),
\end{align} 
where $\matr{P}$ and $\matr{Q}$ are two permutation matrices of size $N\times N$. %In \cite{smc11}, a set of minimal conditions for permutation matrices $\matr{P}$ and $\matr{Q}$ were derived in order to guarantee girth greater than $4,8,$ and $12$.

\item Recall that, by applying Theorem \ref{thm:qcbound} to the base matrix $\base$, we find that any code drawn from the QC sub-ensemble $\xi^{QC}_{\mathbf{B}}(N)$ has minimum distance at most $6$, or equivalently, girth at most $12$. In other words, we cannot exceed a girth of $12$ unless we choose non-circulant permutation matrices $\matr{P}$ and $\matr{Q}$.% In Theorem \ref{th:girth:conditions:2by3}, a set of minimal conditions for $\matr{P}$ and $\matr{Q}$ were given in order to guarantee girth greater than $12$. We will now show that these conditions can be achieved for {circulant-block} permutation matrices $\matr{P}$ and $\matr{Q}$.
%\subsubsection
\end{itemize}
\subsubsection{Pre-lifting a $2\times 3$ protograph} 
\begin{itemize} \item A pre-lifted base matrix $\bpre$ can be written, without loss of generality, as
\begin{align}\bpre=\label{matrix23prelift}
&\begin{bmatrix} \matr{I}_0^m &\matr{I}_0^m &\matr{I}_0^m\\\matr{I}_0^m&\matr{B}_{2,2} & \matr{B}_{2,3}
\end{bmatrix}\in \xi_{\matr{B}}(m).
\end{align} 
%Note that every base matrix $\base ^\prime$ obtained at the pre-lifting step can be re-written in the form of (\ref{matrix1}), and as such, the search space for good covering graphs at the pre-lifting step is reduced to $m!$ permutation choices for $\matr{P}$ and $\matr{Q}$ where $m$ is small. How to choose the covering graph to use at the pre-lifting step will be discussed in more detail later.
%Note that, since base matrix $\base ^\prime$ is lifted from $\base$, it has, without loss of generality, the form of (\ref{matrix1}), and as such, the search space for good covering graphs at the pre-lifting step is reduced to $m!$ permutation choices for $\matr{P}$ and $\matr{Q}$ where $m$ is small. How to choose the covering graph to use at the pre-lifting step will be discussed in more detail later.
Note that, since $\bpre$ is $m$-lifted from $\base$, the search space for good pre-lifted base matrices $\bpre$ consists of at most $m!^2$ combinations of permutation matrices, where $m$ is typically a small integer. %How to choose the permutation matrices at the pre-lifting step will be discussed in more detail in the following.
%\begin{equation}
% \matr{P}=\left[\begin{array}{cc}
%P_1 & 0\\
%0 & P_1
%\end{array}\right]\textrm{ and }
% \matr{\matr{Q}}=\left[\begin{array}{cc}
%0 & \matr{Q}_1\\
%\matr{Q}_2 & 0
%\end{array}\right]
%\end{equation}

\item A QC-LDPC code can now be $r$-lifted from $\bpre$ as
\begin{align*}
\matr{H} &= \hpre \\&=
\left[\begin{array}{ccc}
(\matr{I}_0^m)^{\upmapsto r} & (\matr{I}_0^m)^{\upmapsto r} &(\matr{I}_0^m)^{\upmapsto r} \\
(\matr{I}_0^m)^{\upmapsto r} & (\matr{B}_{2,2})^{\upmapsto r} &(\matr{B}_{2,3})^{\upmapsto r} 
\end{array}\right]\in\xi^{QC}_{\bpre}(r).
\end{align*}
\item Continuing, it can easily be shown (see Appendix \ref{sec:applift}) that by row and column permutations any parity-check matrix $\matr{H}\in\xi^{QC}_{\bpre}(r)$ can be re-written as
\begin{align}\label{23hsimple}
\matr{H}&=\hpre\nonumber\\&=
\left[\begin{array}{ccc}
\matr{I}_0^{mr} & \matr{I}_0^{mr} &\matr{I}_0^{mr} \\
\matr{I}_0^{mr} & (\matr{B}_{2,2})^{\upmapsto r} &(\matr{B}_{2,3})^{\upmapsto r} 
\end{array}\right]\in\xi^{QC}_{\bpre}(r).
\end{align}\end{itemize}
(Note that, as a result of the row and column permutations, the matrices $\matr{B}_{2,2}$ and $\matr{B}_{2,3}$ in  \eqref{23hsimple} are different than the corresponding  matrices in \eqref{matrix23prelift}.) Similar to the simplified representation of $\bpre$ using identity matrices in (\ref{matrix23prelift}), the motivation to write $\matr{H}$ in the form (\ref{23hsimple}) is to simplify the search for suitable circulant permutation matrices at the second lifting step. Instead of searching through $m^6$ combinations of circulants, the search space for good QC-LDPC codes is thus reduced to $m^2$ combinations of circulants. %In general, we use simplified representations of the pre-lifted base matrix $\bpre$ and QC-LDPC matrix $\matr{H}=\hpre$ wherever possible.

For example, consider the pre-lifted base matrix $\matr{B}^{\uparrow 2}$ chosen in (\ref{prelift23}).  
Every parity-check matrix $\matr{H}=\matr{B}^{\uparrow 2\upmapsto r}$ in the ensemble $\xi^{QC}_{\matr{B}^{\uparrow 2}}(r)$ can be written in the form%\vspace{-1mm}
\begin{align}\label{2coverlift}
\matr{H} &= \matr{B}^{\uparrow 2\upmapsto r}= \left[\begin{array}{cc|cc|cc}
\matr{I}_0^r & \matr{0} & \matr{I}_0^r & \matr{0} & \matr{I}_0^r & \matr{0}\\
\matr{0} & \matr{I}_0^r & \matr{0} & \matr{I}_0^r & \matr{0} & \matr{I}_0^r\\\hline
\matr{I}_0^r & \matr{0} & \matr{I}^r_{p_1} & \matr{0} & \matr{0} & \matr{I}^r_{q_1}\\
\matr{0} & \matr{I}_0^r & \matr{0} & \matr{I}^r_{p_2} & \matr{I}^r_{q_2} & \matr{0}\end{array}\right]\nonumber\\ &=
\begin{bmatrix} \matr{I}_0^{2r} &\matr{I}_0^{2r} &\matr{I}_0^{2r}\\\matr{I}_0^{2r}&\matr{P} & \matr{Q}\end{bmatrix},%\vspace{-1mm}
\end{align}
%where permutation matrices $\matr{P}_1,\matr{P}_2,\matr{Q}_1,$ and $\matr{Q}_2$ are of size $r\times r$, and $\matr{P}$ and $\matr{Q}$ are of size $N\times N = 2r \times 2r$. 
 for some $p_1,p_2,q_1,q_2\in[r]$. 
 \begin{itemize} \item By applying Theorem \ref{thm:qcbound}, we find that a code $\cC$ drawn from the QC sub-ensemble $\xi^{QC}_{\matr{B}^{\uparrow 2}}(r)$ with base matrix $\matr{B}^{\uparrow 2}$ from (\ref{prelift23}) has its minimum distance bounded above by $\dmins\leq 10$ (and hence its girth bounded above by $g\leq 20$). \item Note that, by choosing $m=2$, we are forced to use Design Rule 1, because permutations of size $2$ automatically commute. As such, if we hope to achieve increased minimum distance, we must ensure $p_1,p_2,q_1,$ and $q_2$ are chosen such that $\matr{PQ}$ and $\matr{QP}$ do not have an overlapping column.
\end{itemize}
 It can easily be shown that the improvement in minimum distance and girth promised by the application of Theorem \ref{thm:qcbound} can be achieved by codes from $\xi^{QC}_{\bpre}(r)$. For example, 
 \begin{itemize} \item choosing  lifting factor $r=9$ and $(p_1, p_2, q_1, q_2)=(1,2,0,6)$, %respectively,
 %$\matr{P}_1=\matr{I}_1,P_2=\matr{I}_2,\matr{Q}_1=I,$ and $\matr{Q}_2=\matr{I}_6$
 gives a $[54,19,8]$ code with girth $g=16$;
 \item  choosing  lifting factor $r=20$ and $(p_1, p_2, q_1, q_2)=(1,9,0, 4)$, % respectively,
% $\matr{P}_1=\matr{I}_1,\matr{P}_2=\matr{I}_9,\matr{Q}_1=I,$ and $\matr{Q}_2=\matr{I}_{4}$ 
gives a $[120,41,10]$ code with girth $g=20$. 

\end{itemize} For the pre-lifting configuration of (\ref{2coverlift}), we find that $r=9$ and $r=20$ are the smallest possible circulant sizes that enable us to construct codes with girths $16$ and $20$, corresponding to minimum distances $\dmins=8$ and $\dmins=10$, respectively. There are $216$ (resp. $2880$) such codes in the $r=9$ (resp. $r=20$) QC sub-ensembles.

\begin{remark} {\em By choosing $\matr{P}$ and $\matr{Q}$ to be arrays of circulants, or {circulant-block matrices}, rather than just searching for random permutations, we obtain a significant improvement in both the girth and the minimum distance compared to a direct circulant lifting while maintaining the desirable circulant structure facilitating simplified encoding and decoding. Moreover, the search space for good codes is greatly reduced. {In searching for a code with pre-lifting factor $m$ and circulant lifting factor $r$, the circulant-block permutation matrix $\matr{P}$ has $m!\cdot r^m$ choices, or $r^m$ choices after the pre-lifting stage, whereas there are $(mr)!$ choices for a general permutation matrix $\matr{P}$ of size $mr$.} For example, in searching for a code with minimum distance $\dmins=8$ when $m=2$ and $r=9$, {there are $m!\cdot r^m=162$ choices before pre-lifting, or $r^m=81$ choices after pre-lifting, for each of the circulant-block permutation matrices,} whereas there are $mr!=18!\approx 6.4\times 10^{15}$ choices for a random permutation matrix of size $mr=18$. {Note that the number of choices grows quickly with $m$; thus the pre-lifting factor $m$ should be chosen to be small.}}
\end{remark}

\subsubsection{Choosing $m$-fold graph covers for pre-lifting a protograph}
Not all choices of covering graphs are equivalent at the pre-lifting step. For example, the possible choices for the submatrix $[\hspace{0.5mm} \matr{B}_{2,2} \hspace{0.5mm}| \hspace{0.5mm}\matr{B}_{2,3}\hspace{0.5mm}]$ in (\ref{matrix23prelift}) at the pre-lifting step are%\vspace{-1mm}
\begin{eqnarray}
 \left[\begin{array}{cc|cc}\label{2cov1}
1 &0 & 1 & 0\\
0 & 1 & 0 & 1\end{array}\right],\\
\left[\begin{array}{cc|cc}
1 & 0 & 0 & 1\\
0 & 1 & 1 & 0 \end{array}\right], \label{2cov2}\\
 \left[\begin{array}{cc|cc}
0 & 1 & 1 & 0\\
1 & 0 & 0 & 1\end{array}\right], \label{2cov3}\\
\left[\begin{array}{cc|cc}
0 & 1 & 0 & 1\\
1 &0 & 1 & 0\end{array}\right].\hspace{0.5mm}\label{2cov4}
%\vspace{-1mm}
% \left[\begin{array}{cc|cc}\label{2cov1}
%\matr{I}_{p_1}^r & \matr{0} & \matr{I}_{q_1}^r & \matr{0}\\
%\matr{0} & \matr{I}_{p_2}^r & \matr{0} &\matr{I}_{q_2}^r\end{array}\right],\\
%\left[\begin{array}{cc|cc}
%\matr{I}_{p_1}^r & \matr{0} & \matr{0} & \matr{I}_{q_1}^r\\
%\matr{0} & \matr{I}_{p_2}^r  &\matr{I}_{q_2}^r & \matr{0}\end{array}\right],\label{2cov2}\\
% \left[\begin{array}{cc|cc}
%\matr{0} & \matr{I}_{p_1}^r & \matr{I}_{q_1}^r & \matr{0}\\
%\matr{I}_{p_2}^r  & \matr{0} & \matr{0} & \matr{I}_{q_2}^r\end{array}\right],\label{2cov3}\\
%\left[\begin{array}{cc|cc}
%\matr{0} & \matr{I}_{p_1}^r & \matr{0} &\matr{I}_{q_1}^r\\
%\matr{I}_{p_2}^r  & \matr{0} & \matr{I}_{q_2}^r & \matr{0}\end{array}\right].\hspace{0.57mm}\label{2cov4}%\vspace{-1mm}
\end{eqnarray}
Note that choices  \eqref{2cov2},  \eqref{2cov3}, 
and \eqref{2cov4} result in \emph{equivalent} base matrices $\matr{B}^{\uparrow 2}$, \emph{i.e.}, they can be shown to be equal using only elementary row and column operations. Consequently, their lifted ensembles $\xi^{QC}_{\matr{B}^{\uparrow 2}}(r)$ consist of the same set of codes, up to row and column permutations. %Thus, it is necessary to consider only $2$ choices of covering graph for the pre-lifting step with $m=2$, (\ref{2cov1}) and (\ref{2cov2}). 

Applying the bound (\ref{qcbound}) to the pre-lifted configuration (\ref{2cov1}), we find that a code $\cC$ from the QC sub-ensemble $\xi^{QC}_{\matr{B}^{\uparrow 2}}(r)$ has its minimum distance bounded above by $\dmins \leq 12$.
 %\begin{table}[h]
%\begin{center}
%\begin{tabular}{|c|c|c|} 
%\hline  {circulant-block} structure of $P$ and $Q$ & $\dmins(C)\leq$ & $\girth(\matr{H})\leq$\\
%\hline
%(\ref{2cov1}) & $12$ & $24$\\
%(\ref{2cov2}) & $10$ & $20$\\
%(\ref{2cov4}) & $10$ & $20$\\
%\hline
% \end{tabular}
% \end{center}
%\caption{Upper bounds on the minimum Hamming Distance.}\label{tab:nonuniform}
%\end{table}
%$$
%\begin{bmatrix} I &I &I\\I&P_1 &Q_1
%\end{bmatrix}\textrm{ and }\begin{bmatrix} I &I &I\\I&P_2 &Q_2
%\end{bmatrix},
%$$
However, note that the Tanner graph of base matrix $\matr{B}^{\uparrow 2}$ corresponding to (\ref{2cov1}) consists of two disconnected copies of the original protograph. It follows that any lifted parity-check matrix contains the two following disjoint substructures: 
$$
\begin{bmatrix} \matr{I}_0^r &\matr{I}_0^r &\matr{I}_0^r\\\matr{I}_0^r&\matr{I}_{p_1}^r &\matr{I}_{q_1}^r
\end{bmatrix}\textrm{ and }\begin{bmatrix} \matr{I}_0^r &\matr{I}_0^r &\matr{I}_0^r\\\matr{I}_0^r&\matr{I}_{p_2}^r &\matr{I}_{q_2}^r
\end{bmatrix},%\vspace{-1mm}
$$ and consequently its minimum distance and girth are bounded above by $\dmins\leq 6$ and $g\leq 12$, respectively. Thus, in terms of maximizing minimum distance and girth, {the pre-lifting configuration }(\ref{2cov2}){, or the equivalent pre-lifting configurations }\eqref{2cov3}{ or }\eqref{2cov4}{, should be chosen}.

\subsubsection{Larger degrees of pre-lifting} Intuitively, the larger we make the pre-lifting factor $m$ for a fixed block length $n$, the more `random-like' the QC sub-ensemble $\xi^{QC}_{\bpre}(r)$ becomes and, as a consequence, we would expect the maximum achievable minimum distance to increase. We have seen that, for $m=2$, the maximum achievable minimum distance of a circulant-based lifting increased from $\dmins\leq 6$ to $\dmins\leq 10$, and correspondingly, the maximum achievable girth increased from $12$ to $20$. In the remainder of this section, we describe how the minimum distance and girth are affected by increasing the pre-lifting factor to values of $m\geq 3$. Note that, for $m\geq 3$, the permutation matrices do not necessarily commute with one another, so both Design Rules 1 and 2 may be used.

We employ the sieve principle (start with all possibilities, perform a test, remove candidates that fail the test, and repeat until we can no longer separate the candidates) in order to find a good covering graph to use at the pre-lifting step. Note that every $3$-cover can be written in the form of (\ref{matrix23prelift}), and as such, there are $m!^2=3!^2=36$ covering graphs to consider for $m=3$. Of these $3$-covers, we find that many are equivalent. In fact, after removing (or sieving out) equivalent graphs, we are left with only five choices. Of these choices, if any contain disjoint sub-graphs of a smaller covering graph ($m=1$ or $m=2$ in this case), then the minimum distance cannot exceed the corresponding bound calculated for the sub-graph. For a $3$-cover, there are two such sub-graphs; either there are three copies of the $1$-cover (3 disjoint copies of the original protograph), or the lifted graph consists of both a $1$-cover and a $2$-cover (a copy of the original protograph and a disjoint $2$-cover). In both cases, a code $\cC$ drawn from the QC sub-ensemble has its minimum  distance bounded above by $\dmins\leq 6$ as a result of the substructure associated with the $1$-cover. For example, the only configuration of $[\hspace{0.5mm} \matr{B}_{2,2} \hspace{0.5mm}| \hspace{0.5mm}\matr{B}_{2,3}\hspace{0.5mm}]$  that results in three copies of the $1$-cover is when both  $\matr{B}_{2,2}$ and $ \matr{B}_{2,3}$ are identity matrices, \emph{i.e.}, the circulants in the lifted {circulant-block} matrix occur only on the leading diagonal.
%\begin{equation}
% \left[\begin{array}{ccc|ccc}\label{3cov1}
%\matr{P}_1 & \matr{0} & \matr{0} & \matr{Q}_1 & \matr{0}& \matr{0}\\
%\matr{0} & \matr{P}_2 & \matr{0} & \matr{0} & \matr{Q}_2& \matr{0}\\
%\matr{0} & \matr{0} & \matr{P}_3 & \matr{0} & \matr{0} & \matr{Q}_3\end{array}\right].
%\end{equation}
There are nine (equivalent) occurrences of the second limiting substructure consisting of both a $1$-cover and a $2$-cover. One such example is the substructure
\begin{equation}
[\>\matr{B}_{2,2}\>|\>\matr{B}_{2,3}\>] %= \left[\begin{array}{cc}\matr{B}_{2,2} &\matr{B}_{2,3}\end{array}\right] 
= \left[\begin{array}{ccc|ccc}\label{3cov1}
%\matr{I}_{p_1}^r & \matr{0} & \matr{0} & \matr{0} &\matr{I}_{q_1}^r& \matr{0}\\
%\matr{0} & \matr{I}_{p_2}^r & \matr{0} & \matr{I}_{q_2}^r & \matr{0}& \matr{0}\\
%\matr{0} & \matr{0} & \matr{I}_{p_3}^r & \matr{0} & \matr{0} & \matr{I}_{q_3}^r\end{array}\right],%\vspace{-2mm}
1 & 0 & 0 & 0 & 1 & 0\\
0 & 1 & 0 & 1 & 0 & 0\\
0 & 0 & 1 & 0 & 0 & 1\end{array}\right],%\vspace{-2mm}
\end{equation}
which again results in any code $\cC$ drawn from $\xi^{QC}_{\matr{B}^{\uparrow 3}}(r)$ having its minimum distance bounded above by $\dmins\leq 6$ for arbitrarily large circulant size $r$.

Note that applying (\ref{qcbound}) to base matrices containing these two harmful substructures gives the loose upper bounds $\dmins\leq 24$ and $\dmins\leq 12$, respectively, and so it is necessary to remove these candidates before before proceeding with the code construction. After removing the equivalent covering graphs and those containing disjoint subgraphs, we are left with three candidates for the pre-lifted base matrix $\matr{B}^{\uparrow 3}$. Applying (\ref{qcbound}) to the remaining choices results in one candidate that bounds the minimum distance of circulant-based codes drawn from the ensemble by $\dmins\leq 10$ and two (non-equivalent) candidates with bound $\dmins\leq 12$. Note that $\dmins\leq 10$ is achievable by a $2$-cover, so this choice is removed, leaving only two remaining choices for the pre-lifted graph. One of the remaining choices is the $3$-cover with the following sub-matrix (before and after the second lifting step)
\begin{align}
[\>\matr{B}_{2,2} \>|\>\matr{B}_{2,3}\>] =& \left[\begin{array}{ccc|ccc}
%\matr{P}_1 & \matr{0} & \matr{0} & \matr{0} & \matr{Q}_1& \matr{0}\\
%\matr{0} & \matr{P}_2 & \matr{0} & \matr{0} & \matr{0}& \matr{Q}_2\\
%\matr{0} & \matr{0} & \matr{P}_3 & \matr{Q}_3 & \matr{0} & \matr{0}
1& 0 & 0& 0&1  & 0\\
0&1 & 0& 0& 0 &  1\\
0& 0&1  &  1 & 0 & 0
\end{array}\right]
 \leadsto  \nonumber \\  
 [\>\matr{H}_{2,2} \>|\>\matr{H}_{2,3}\>] =& 
 \left[\begin{array}{ccc|ccc}\label{3cov2}
%\matr{P}_1 & \matr{0} & \matr{0} & \matr{0} & \matr{Q}_1& \matr{0}\\
%\matr{0} & \matr{P}_2 & \matr{0} & \matr{0} & \matr{0}& \matr{Q}_2\\
%\matr{0} & \matr{0} & \matr{P}_3 & \matr{Q}_3 & \matr{0} & \matr{0}
\matr{I}_{p_1}^r & \matr{0} & \matr{0} & \matr{0} & \matr{I}_{q_1}^r  & \matr{0}\\
\matr{0} & \matr{I}_{p_2}^r  & \matr{0} & \matr{0} & \matr{0} &  \matr{I}_{q_2}^r\\
\matr{0} & \matr{0} & \matr{I}_{p_3}^r  &  \matr{I}_{q_3}^r & \matr{0} & \matr{0}
\end{array}\right].
\end{align}
Note that this choice of pre-lifting again forces us to use Design Rule 1, since the permutation matrices in $\matr{B}^{\uparrow 3}$ all commute with one another. Choosing circulants $\matr{I}_{p_1}^{46},\matr{I}_{p_2}^{46},\matr{I}_{p_3}^{46},\matr{I}_{q_1}^{46},\matr{I}_{q_2}^{46}$, and $\matr{I}_{q_3}^{46}$ as $\matr{I}_1^{46},\matr{I}_5^{46},\matr{I}_{25}^{46},\matr{I}_4^{46},\matr{I}_7^{46},$ and $\matr{I}_{28}^{46}$, respectively, results in a code $\cC$ with minimum distance $\dmins=12$ and girth $g=24$, and we see that the corresponding bound can be achieved.

The procedure can be repeated for $m\geq 4$. Applying the sieve technique to the $4!^2$ candidate covering graphs for $m=4$, we are left with five candidates for which $\dmins\leq 14$. Codes achieving a minimum distance equal to $14$ can be constructed, so we see again that the bound can be achieved. Table \ref{tab:2covers} summarizes the results we have obtained as a result of pre-lifting the $2\times 3$ all-ones base matrix $\base$.
 \begin{table}[h]
 \caption{Largest achievable values of minimum distance and girth for a $(2,3)$-regular base matrix given a particular pre-lifting factor $m$.}\label{tab:2covers}%\vspace{-3mm}
\begin{center}
\begin{tabular}{|c|c|c|} 
\hline pre-lifting factor $m$  & $\dmins$ & girth\\
\hline
$1$ & $6$ & $12$\\
$2$ & $10$ & $20$\\
$3$ & $12$ & $24$\\
$4$ & $14$ & $28$\\

\hline
 \end{tabular}
 \end{center}

\end{table}%\vspace{-1mm}

Note that the minimum distance grows slowly in this example, but this is expected for $(2,3)$-regular codes (see \cite{gal62}). It does, however, demonstrate that the minimum distance and girth can be improved by pre-lifting the protograph. In the next section we will obtain larger improvements by considering a protograph with increased node degrees. %\vspace{-3mm}

\subsubsection{Discussion}
The pre-lifting technique described above is a simple but effective way to improve the performance of QC-LDPC codes. In fact, many existing QC-LDPC codes in the literature can be viewed as pre-lifted codes. In this subsection, we compare some of our constructions to known optimal and close to optimal codes (in the sense of minimal block length for a given $d_\mathrm{min}$). For example, we know that a code can be constructed with $d_\mathrm{min} = 6$ by directly lifting $\matr{H}$ from $\matr{B}$, \emph{i.e.}, $m=1$. In fact, the minimal length $[21,8,6]$ code meeting this criteria {can be viewed as a (degenerate) pre-lifted graph} with  $m=1$ and $r=7$, where the parity-check matrix\begin{equation*}
\mathbf{H} = \left[\begin{array}{ccc}
\mathbf{I}_0^7 & \mathbf{I}_0^7 & \mathbf{I}_0^7\\
\mathbf{I}_0^7 & \mathbf{I}_4^7 & \mathbf{I}_6^7
\end{array}\right]
\end{equation*}
determines the Heawood graph \cite{bks09}. To obtain $d_\mathrm{min} = 8$, we see from Table~\ref{tab:2covers} that it is necessary to increase the pre-lifting factor to at least $m=2$. It is known that the shortest possible (optimal) $(2,3)$-regular code has parameters $[45,16,8]$ \cite{bks09}, which is not too far from the pre-lifted $[54, 19,8]$ code constructed in Section~\ref{sec:2x3}-1 with $m=2$ and $r=9$ (which was obtained with no particular effort to minimize block length). In addition, we note that the optimal  $[45,16,8]$ code \cite{bks09} can be viewed as a pre-lifted code from $\matr{B}$ with $m=3$ and $r=5$, where the parity-check matrix is:
\begin{equation*}
\mathbf{H} = \left[\begin{array}{ccc|ccc|ccc}
\mathbf{I}_0^5 & \mathbf{0} & \mathbf{0} & \mathbf{I}_0^5 & \mathbf{0} & \mathbf{0} & \mathbf{I}_0^5 & \mathbf{0} & \mathbf{0}\\
\mathbf{0} & \mathbf{I}_0^5 & \mathbf{0} & \mathbf{0} & \mathbf{I}_0^5 & \mathbf{0} & \mathbf{0} & \mathbf{I}_0^5 & \mathbf{0}\\
\mathbf{0} & \mathbf{0} & \mathbf{I}_0^5 & \mathbf{0} & \mathbf{0} & \mathbf{I}_0^5 & \mathbf{0} & \mathbf{0} & \mathbf{I}_0^5\\\hline
\mathbf{I}_0^5 & \mathbf{0} & \mathbf{0} & \mathbf{0} &\mathbf{0}  & \mathbf{I}_0^5& \mathbf{0} & \mathbf{I}_1^5 & \mathbf{0}\\
\mathbf{0} & \mathbf{I}_0^5 & \mathbf{0} & \mathbf{I}_0^5 & \mathbf{0} & \mathbf{0} &\mathbf{I}_3^5& \mathbf{0} & \mathbf{0}\\
\mathbf{0} & \mathbf{0} & \mathbf{I}_0^5 & \mathbf{0}  & \mathbf{I}_4^5 & \mathbf{0} & \mathbf{0} & \mathbf{0} & \mathbf{I}_1^5\\
\end{array}\right].
\end{equation*}
Consequently, it is clear that, for a given desired block length $n=n_vmr$ and required $d_\mathrm{min}$, it is an interesting challenge to choose the correct degree of pre-lifting $m$. Generally, to reduce complexity, we choose $m$ as small as possible to achieve a desired $d_\mathrm{min}$; however, as we see in this example, it is possible that such a $d_\mathrm{min}$ can be obtained with a shorter overall block length by choosing a larger $m$ and smaller $r$.

Finally, we point out that for $d_\mathrm{min}  = 10$ the  $[120, 41,10]$ pre-lifted code constructed in Section~\ref{sec:2x3}-1 is also close to the optimal $[105,36,10]$ code based on the Balaban graph \cite{bks09} and to the near-optimal $[108,37,10]$ code presented in \cite{bks09}. The fact that the pre-lifted codes, which were constructed for demonstration purposes without any particular effort to minimize block length, are close to the lower bounds on block length for a given $d_\mathrm{min}$ demonstrates the efficiency of this method. Useful references and short tables of near-optimal $(2, K)$-regular LDPC codes can be found in \cite{bks09}.

%%%%%%%%%%%%%%%%%%%%%%%%%%%%%%%%%%%%%%%%%%%%%%%%%%%
\subsection{Pre-lifted QC structures for a $3 \times 4$ base matrix}\label{sec:3x4}
%%%%%%%%%%%%%%%%%%%%%%%%%%%%%%%%%%%%%%%%%%%%%%%%%%%
%\begin{example}\label{ex:34prelift}
Consider the $(3,4)$-regular protograph-based ensemble defined by the all-ones base matrix $\matr{B}$ of size $3\times 4$. 
\begin{itemize} \item We can assume, without loss of generality, that any parity-check matrix derived from $\matr{B}$ can be written in the form of \eqref{matrix:general1}, \emph{i.e.},
 \begin{align}\label{matrix3by4}\matr{H}=\matr{B}^{\uparrow N}=\begin{bmatrix} \matr{I}_{0}^N &\matr{I}_{0}^N &\matr{I}_{0}^N&\matr{I}_{0}^N\\ \matr{I}_{0}^N& \matr{P}&\matr{R}&\matr{T}\\ \matr{I}_{0}^N& \matr{Q}&\matr{S}&\matr{U}
\end{bmatrix} \in \xi_{\mathbf{B}}(N),
\end{align}
 \noindent where $\matr{P},\matr{Q},\matr{R},\matr{S},\matr{T}$, and $\matr{U}$ are $N\times N$ permutation matrices. \item We can also assume, without loss of generality, that a pre-lifted base matrix $\bpre$ has the form
\begin{equation}\label{34preliftgeneral}
 \bpre=\left[\begin{array}{cccc}
 \matr{I}_{0}^m & \matr{I}_{0}^m & \matr{I}_{0}^m & \matr{I}_{0}^m\\
 \matr{I}_{0}^m & \matr{B}_{2,2} & \matr{B}_{2,3}& \matr{B}_{2,4}\\
 \matr{I}_{0}^m &  \matr{B}_{3,2} & \matr{B}_{3,3}& \matr{B}_{3,4}

\end{array}\right]\in\xi_{\mathbf{B}}(m),
\end{equation}
where $\matr{B}_{2,2}$, $\matr{B}_{3,2}$, $\matr{B}_{2,3}$, $\matr{B}_{3,3}$, $\matr{B}_{2,4}$, and $\matr{B}_{3,4}$ are $m\times m$ permutation matrices. 
\item Finally, any parity-check matrix $\matr{H}\in\xi_{\bpre}^{QC}(r)$ can be written as
\begin{align}\label{34hgeneral}
\matr{H} &= \hpre\nonumber\\&=\left[\begin{array}{cccc}
 \matr{I}_{0}^{mr} & \matr{I}_{0}^{mr} & \matr{I}_{0}^{mr} & \matr{I}_{0}^{mr}\\
 \matr{I}_{0}^{mr} & (\matr{B}_{2,2})^{\upmapsto r} & (\matr{B}_{2,3})^{\upmapsto r}& (\matr{B}_{2,4})^{\upmapsto r}\\
 \matr{I}_{0}^{mr} &(\matr{B}_{3,2})^{\upmapsto r} & (\matr{B}_{3,3})^{\upmapsto r}& (\matr{B}_{3,4})^{\upmapsto r}\end{array}\right]
\end{align}
after row and column permutations. Note that (\ref{34hgeneral}) is in the form of  (\ref{matrix3by4}), where $N=mr$ and $\matr{P}$, $\matr{Q}$, $\matr{R}$, $\matr{S}$, $\matr{T}$, and $\matr{U}$ are {circulant-block} permutation matrices. Note also that, as a result of the row and column permutations, the matrices $\matr{B}_{i,j}$ in  \eqref{34hgeneral} are generally different than the corresponding  matrices in \eqref{34preliftgeneral}.
\end{itemize}

Using the technique presented in \cite{smc11}, we determine that, for any parity-check matrix in the form of  (\ref{matrix3by4}), we can ensure 
\begin{itemize} \item $g\geq 6$ if all of the $18$ matrices in the following set do not have a fixed column:
\begin{align} \label{g6}
&\{\matr{P},\matr{Q},\matr{R},\matr{S},\matr{T},\matr{U},\notag\\
&\matr{PQ}^\tr\hspace{-0.5mm},\matr{PR}^\tr\hspace{-0.5mm}, \matr{PT}^\tr\hspace{-0.5mm},\matr{QS}^\tr\hspace{-0.5mm},  \matr{QU}^\tr\hspace{-0.5mm},\matr{RS}^\tr\hspace{-0.5mm},\matr{RT}^\tr\hspace{-0.5mm},  \matr{SU}^\tr\hspace{-0.5mm}, \matr{TU}^\tr\hspace{-0.5mm},\notag\\
& \matr{PQ}^\tr \matr{SR}^\tr\hspace{-0.5mm}, \matr{PQ}^\tr \matr{UT}^\tr\hspace{-0.5mm}, \matr{RS}^\tr \matr{UT}^\tr\hspace{-0.5mm}\};
\end{align}

 \item $g\geq 8$ if all of the $42$ matrices in the following set do not have a fixed column:
\begin{align}\label{g8}
&\{\matr{P},\matr{Q},\matr{R},\matr{S},\matr{T},\matr{U},\notag\\&\matr{PQ}^\tr\hspace{-0.5mm},\matr{PR}^\tr\hspace{-0.5mm}, \matr{PS}^\tr\hspace{-0.5mm},\matr{PT}^\tr\hspace{-0.5mm}, \matr{PU}^\tr\hspace{-0.5mm},\matr{QR}^\tr\hspace{-0.5mm},\notag
\matr{QS}^\tr\hspace{-0.5mm}, \matr{QT}^\tr\hspace{-0.5mm}, \matr{QU}^\tr\hspace{-0.5mm},\\&\matr{RS}^\tr\hspace{-0.5mm},\matr{RT}^\tr\hspace{-0.5mm},\matr{RU}^\tr\hspace{-0.5mm},  \matr{ST}^\tr\hspace{-0.5mm},\matr{SU}^\tr\hspace{-0.5mm}, \matr{TU}^\tr\hspace{-0.5mm},\notag\\
&\matr{PSR}^\tr\hspace{-0.5mm}, \matr{PUT}^\tr\hspace{-0.5mm},\matr{RUT}^\tr\hspace{-0.5mm}, \matr{TS} \matr{R}^\tr ,\matr{RQ} \matr{P}^\tr\hspace{-0.5mm},\matr{TQ} \matr{P}^\tr ,\notag\\
&\matr{RS}^\tr \matr{Q},\matr{RS}^\tr \matr{U}, \matr{TU}^\tr \matr{Q},\matr{TU}^\tr \matr{S}, \matr{PQ}^\tr \matr{S},\matr{PQ}^\tr \matr{U}, \notag\\
& \matr{PQ}^\tr \matr{SR}^\tr\hspace{-0.5mm}, \matr{PQ}^\tr \matr{ST}^\tr ,\matr{PQ}^\tr  \matr{UR}^\tr\hspace{-0.5mm}, \matr{PQ}^\tr \matr{UT}^\tr\hspace{-0.5mm},\matr{PS}^\tr  \matr{UT}^\tr\hspace{-0.5mm},\notag\\
& \matr{RQ}^\tr \matr{UT}^\tr\hspace{-0.5mm},  \matr{RS}^\tr \matr{QT}^\tr\hspace{-0.5mm},\matr{RS}^\tr \matr{UP}^\tr , \matr{RS}^\tr \matr{UT}^\tr\}.\end{align}

 \end{itemize}

Following the same process, additional conditions can be used to guarantee even larger girths. We will see later that, by pre-lifting $\matr{B}$ to $\bpre$, the number of such conditions that must be checked in order to achieve girth $g$ for a derived matrix $\matr{H}=\hpre\in\xi_{\bpre}(r)$ can be significantly less than for a general lifted matrix $\matr{H}=\matr{B}^{\uparrow mr}\in\xi_{\mathbf{B}}(mr)$. Moreover, we can use the circulant-based structure and corresponding modular arithmetic to reduce the complexity of evaluating the conditions and searching for suitable permutation matrices.

Recall that, if we take a direct circulant-based lifting of $\matr{B}$, the existence of a $4$- or $6$-cycle in the Tanner graph automatically implies a codeword of weight less than the upper bound $\dmins\leq 24$, so a minimum girth of $8$ is required to achieve the bound. The $[124,33,24]$ QC Tanner code defined in (\ref{tannercode}) is an example of a code achieving the upper  bound with girth $g=8$.  In the remainder of this section we show that, by pre-lifting the $3\times 4$ all-ones base matrix $\base$, we can construct circulant-based codes with minimum distance exceeding the upper bound $\dmins\leq 24$ for QC codes drawn from $\xi^{QC}_{\mathbf{B}}(N)$, even if a $6$-cycle exists in the graph. Moreover, we observe further improvements by ensuring a girth larger than $6$.

From (\ref{34preliftgeneral}), there are $m^6=64$ possible $2$-covers of $\base$ that can be considered as candidates $\matr{B}^{\uparrow 2}$ for the pre-lifting step. After removing equivalent covering graphs (the $2$-covers that are equal after re-labeling the vertices) there are five candidates left. Note that the only harmful substructure to avoid in a $2$-cover is the single occurrence of two disjoint $1$-covers. This can only occur if $\matr{B}_{i,j}=\matr{I}_0^m$ for all $(i,j)\in \{2,3\}\times\{2,3,4\}$. Any code $\cC$ drawn from this QC sub-ensemble $\xi^{QC}_{\matr{B}^{\uparrow 2}}(r)$ will have minimum distance bounded above by $\dmins\leq 24$ for arbitrarily large $r$. After removing this $2$-cover, we have only four remaining candidates. Of these candidates, two give $\dmins\leq 120$ and two give $\dmins\leq 116$, both significantly larger than the bound for the $1$-cover, $\dmins\leq 24$. 
\begin{example}\label{ex:34liftchoice1} Consider the following $2$-cover of $\base$
\begin{equation}\label{342cov}
\matr{B}^{\uparrow 2}= \left[\begin{array}{cc|cc|cc|cc}
1 & 0 & 1 & 0 & 1 & 0 & 1 & 0\\
0 & 1 & 0 & 1 & 0 & 1 & 0 & 1\\\hline
1 & 0 & 1 & 0 & 0 & 1 & 0 & 1\\
0 & 1 & 0 & 1 & 1 & 0 & 1 & 0\\\hline
1 & 0 & 0 & 1 & 1 & 0 & 1 & 0\\
0 & 1 & 1 & 0 & 0 & 1 & 0 & 1\end{array}\right],
\end{equation}
from which, without loss of generality, any lifted code in the ensemble $\xi_{\bpre}^{QC}(r)$ has the parity-check matrix
\begin{align}
\matr{H}= \matr{B}^{\uparrow 2 \upmapsto r}=\nonumber
&\left[\begin{array}{cc|cc|cc|cc}\label{342cov2}
\matr{I}_0^r & \matr{0} & \matr{I}_0^r & \matr{0} & \matr{I}_0^r & \matr{0} & \matr{I}_0^r & \matr{0}\\
\matr{0} & \matr{I}_0^r & \matr{0} & \matr{I}_0^r & \matr{0} & \matr{I}_0^r & \matr{0} & \matr{I}_0^r\\\hline
\matr{I}_0^r & \matr{0} & \matr{I}_{p_1}^r & \matr{0} & \matr{0} & \matr{I}_{r_1}^r & \matr{0} &  \matr{I}_{t_1}^r\\
\matr{0} & \matr{I}_0^r & \matr{0} &  \matr{I}_{p_2}^r &  \matr{I}_{r_2}^r & \matr{0} &  \matr{I}_{t_2}^r & \matr{0}\\\hline
\matr{I}_0^r & \matr{0} & \matr{0} &  \matr{I}_{q_1}^r &  \matr{I}_{s_1}^r & \matr{0} &  \matr{I}_{u_1}^r & \matr{0}\\
\matr{0} & \matr{I}_0^r &  \matr{I}_{q_2}^r & \matr{0} & \matr{0} & \matr{I}_{s_2}^r & \matr{0} &  \matr{I}_{u_2}^r\end{array}\right]\\=&
\begin{bmatrix} \matr{I}_{0}^{2r} &\matr{I}_{0}^{2r}  &\matr{I}_{0}^{2r} &\matr{I}_{0}^{2r} \\ \matr{I}_{0}^{2r} & \matr{P}&\matr{R}&\matr{T}\\ \matr{I}_{0}^{2r} & \matr{Q}&\matr{S}&\matr{U}
\end{bmatrix}.%\vspace{-0.5mm}
\end{align}
%where $\matr{P}_i,\matr{Q}_i,\matr{R}_i,\matr{S}_i,\matr{T}_i,$ and $\matr{U}_i$, $i=1,2$, are permutation matrices of size $r\times r$. 
Using (\ref{qcbound}), we find that codes drawn from  $\xi^{QC}_{\matr{B}^{\uparrow 2}}(r)$ have minimum distance bounded above by $\dmins\leq 116$. Note that, because we have a small lifting factor $m=2$, all of the permutations in $\matr{B}^{\uparrow 2}$ are circulant, \emph{i.e.}, every pair of submatrices commute. In this case we must use Design Rule 1 and make sure the circulants chosen at the second step allow the minimum distance to exceed $24$. 

We also wish to ensure that the Tanner graph has an acceptable girth. Recall that if the $18$ matrices in (\ref{g6}) do not have a fixed column, then $g\geq 6$, and if the $42$ matrices in (\ref{g8}) do not have a fixed column, then $g\geq 8$. By applying Lemma \ref{thm:girthlemma} with the {circulant-block} permutation matrices from (\ref{342cov}), we find that the number of conditions from (\ref{g6}) and (\ref{g8}) that we need to check is reduced to $8$ and $20$, respectively. As an example, one surviving condition is that 
\begin{align*}
&\matr{RT}^\tr \nonumber= \left[\hspace{-0.5mm}\begin{array}{cc} 
\matr{0} & \matr{I}_{r_1}^r\\
\matr{I}_{r_1}^r & \matr{0}
\end{array}\hspace{-0.5mm}\right]
\left[\hspace{-0.5mm}\begin{array}{cc} 
\matr{0} & \matr{I}_{t_1}^r\\
\matr{I}_{t_2}^r & \matr{0}
\end{array}\hspace{-0.5mm}\right]^\tr \\&=
 \left[\hspace{-0.5mm}\begin{array}{cc} 
\matr{0} & \matr{I}_{r_1}^r\\
\matr{I}_{r_1}^r & \matr{0}
\end{array}\hspace{-0.5mm}\right]
\left[\hspace{-0.5mm}\begin{array}{cc} 
\matr{0} & \matr{I}_{r-t_2}^r\\
\matr{I}_{r-t_1}^r & \matr{0}
\end{array}\hspace{-0.5mm}\right] =
\left[\hspace{-0.5mm}\begin{array}{cc} 
\matr{I}_{r_1+r -t_1}^r & \matr{0}\\
 \matr{0} & \matr{I}_{r_1+r-t_2}^r 
\end{array}\hspace{-0.5mm}\right]
\end{align*}
should not have a fixed column.  This can be achieved simply by ensuring that $r_i+r -t_i \not\equiv 0 \mod r$, for $i=1,2$.

\begin{itemize} \item Choosing the shift parameters \\%$\matr{P}_1, \matr{P}_2,\matr{Q}_1,\matr{Q}_2,\matr{R}_1,\matr{R}_2,\matr{S}_1,\matr{S}_2,\matr{T}_1,\matr{T}_2,\matr{U}_1$, and $\matr{U}_2$ as circulant matrices $\matr{I}_1, \matr{I}_5,\matr{I}_7,\matr{I}_3,\matr{I}_2,\matr{I}_{10},\matr{I}_{14},\matr{I}_6,\matr{I}_4,\matr{I}_{20},\matr{I}_{28},$ and $\matr{I}_9$
$$(p_1,p_2,q_1,q_2,r_1, r_2,s_1,s_2,t_1,t_2,u_1, u_2)=$$ $$( 1,5,7,3,2,10,14,6,4,20,28, 9),$$ results in {circulant-block} permutation matrices $\matr{P}$, $\matr{Q},$ $\matr{R},\matr{S},\matr{T}, \matr{U}$ that give $g\geq 6$ for $r\geq 31$. For $r=31$, we find that $\dmins=36$ and $g=6$.
\item  By increasing the circulant size to $r=41$, we find that $g=6$ and determine that the minimum  distance is bounded by $38 \leq \dmins \leq 48$ using MAGMA \cite{bcp97}.\footnote{Due to the computational complexity, we are not able to determine the minimum distance of this example exactly. However, we conjecture that it is, in fact, equal or close to the upper bound based on the results obtained for smaller values of $r$ and the significant search time without finding any codewords of weight less than $48$.}  
\item  Recall that direct circulant liftings of $\base$ have minimum distance bounded above by $\dmins\leq 24$ for arbitrarily large circulant size, and a cycle of length $6$ implies $\dmins<24$ (see \cite{sv12}). \end{itemize}\end{example}\hfill $\Box$

\begin{example}\label{ex:34liftchoice2} Choosing the shift parameters in (\ref{342cov2}) to be $$(p_1,p_2,q_1,q_2,r_1, r_2,s_1,s_2,t_1,t_2,u_1, u_2)=
$$  $$(1,5,7,7,10,10,11,11,13,13,2,4),$$  gives $g>6$ for $r\geq 20$. In fact, even for $r=17$, we obtain a $[136,36,26]$ code  with $g=8$. By increasing the circulant size to $r=49$, the code has $g=10$ and we can determine that the minimum  distance is bounded by $32 \leq \dmins \leq 56$ using MAGMA. \end{example}\hfill $\Box$

In this section, we have applied the techniques of pre-lifting to a $(3,4)$-regular protograph. We observed a large improvement in the minimum  distance of QC-LDPC codes lifted from a $2$-cover and we expect further improvement for larger pre-lifting factors $m$. 

%%%%%%%%%%%%%%%%%%%%%%%%%%%%%%%%%%%%%%%%%%%%%%%%%%%
\subsection{Irregular protographs}\label{sec:ireg}
%%%%%%%%%%%%%%%%%%%%%%%%%%%%%%%%%%%%%%%%%%%%%%%%%%%
Luby et al. showed that the performance of LDPC codes can be improved significantly by introducing irregularity into the code graphs \cite{lmss01}. So far in this section we have only considered regular all-ones base matrices, but irregularities can easily be introduced by removing edges of the protograph. This technique, called \emph{masking}, was introduced in \cite{xcd+07} to construct good irregular LDPC codes from arrays of circulants. Masking involves replacing a number of the $N\times N$ permutation matrices with the $N\times N$ all-zero matrix. In particular, masking removes cycles in the graph and can increase the girth.
\begin{example} Consider the masked $(3,4)$-regular base matrix $\matr{B}$ and its corresponding $N$-lifted parity-check matrix  $\matr{H}=\matr{B}^{\uparrow N}$
\begin{align}\label{matrix3by4irreg}
\matr{B}=\left[\begin{array}{cccc} 
1 &1&1&1\\ 1 &0&1&1\\ 1 &1&1&0\\
\end{array}\right] \textrm{ and }\matr{H}=\left[\begin{array}{cccc} 
\matr{I}_0^N &\matr{I}_0^N &\matr{I}_0^N&\matr{I}_0^N\\ \matr{I}_0^N& \matr{0}&\matr{R}&\matr{T}\\ \matr{I}_0^N& \matr{Q}&\matr{S}&\matr{0}
\end{array}\right].
\end{align}
Recall that, for the unmasked ensemble with parity-check matrix given in (\ref{matrix3by4}), in order to achieve girth $g\geq 8$ it was required that each of the $42$ permutation matrices given in (\ref{g8}) should have no fixed columns. For the masked version of $\matr{H}$ in the form of (\ref{matrix3by4irreg}), the number of permutation matrices  that must be checked is reduced to these $13$:
\begin{align}\label{irreggirth}\nonumber
\{ &\matr{Q},\matr{R},\matr{S},\matr{T}, \matr{Q}\matr{S}^\tr, \matr{R}\matr{S}^\tr,  \matr{R}\matr{T}^\tr, \\&
\matr{Q}\matr{R}^\tr,\matr{Q}\matr{T}^\tr,  \matr{S}\matr{T}^\tr,\nonumber \matr{R}\matr{S}^\tr \matr{Q},
\matr{T}\matr{S} \matr{R}^\tr, \matr{R}\matr{S}^\tr \matr{Q}\matr{T}^\tr\},
\end{align}
%\begin{align}\label{irreggirth}\nonumber
%\{ &\matr{Q},\matr{R},\matr{S},\matr{T},\matr{Q}\matr{S}^\tr, \matr{R}\matr{T}^\tr, \matr{R}\matr{S}^\tr,  \matr{R}\matr{Q}^\tr, \matr{T}\matr{Q}^\tr, \\&\matr{T}\matr{S}^\tr,\matr{R}\matr{S}^\tr \matr{Q},
%\matr{R}\matr{S}^\tr \matr{T}^\tr, \matr{R}\matr{S}^\tr \matr{Q}\matr{T}^\tr\},
%\end{align}
where only the first $7$ should have no fixed columns to ensure $g\geq 6$. Moreover, suppose $\matr{B}$ is pre-lifted to $\matr{B}^{\uparrow 2}$ using the permutation matrices from (\ref{342cov}). Then the number of permutation matrices in $\matr{H}=\matr{B}^{\uparrow 2 \upmapsto r}$ that must be checked is further reduced to these $6$:
\begin{align*}\{&\matr{S},  \matr{R}\matr{T}^\tr, %\\&
\matr{Q}\matr{R}^\tr,\matr{Q}\matr{T}^\tr,  \matr{R}\matr{S}^\tr \matr{Q},
\matr{T}\matr{S}\matr{R}^\tr\},\end{align*}
% \begin{equation*}\{ \matr{S}, \matr{R}\matr{T}^\tr,  \matr{R}\matr{Q}^\tr, \matr{T}\matr{Q}^\tr,\matr{R}\matr{S}^\tr \matr{Q},
%\matr{R}\matr{S}^\tr \matr{T}^\tr\},\end{equation*} 
with only the first two matrices needing to be checked to ensure $g\geq 6$.

Note that, while masking can improve the cycle properties of a Tanner graph, it often has a negative effect on minimum distance. \begin{itemize}\item For this example, we find that the upper bound on distance for any QC-LDPC code in $\xi^{QC}_{\matr{B}}(N)$  with parity-check matrix $\matr{H}=\matr{B}^{\upmapsto N}$ is reduced to $\dmins \leq 14$ (recall that for the unmasked case $\dmins \leq 24$). \item This dramatic decrease in minimum distance is likely a result of the large number of weight $2$ columns in $\matr{H}$, and in this case pre-lifting is even more important.{\footnote{{It is well known that the minimum distance properties of both unstructured} \cite{dru06} {and protograph-based} \cite{ddja09} {code ensembles are sensitive to the number of degree two variable nodes in the code graph. In the case of structured code ensembles, such as protograph-based code ensembles, the connectivity of the degree two variable nodes is also important (see, \emph{e.g.},} \cite{ddja09}{).}}} \item We find that by pre-lifting the masked base matrix, the upper bound on minimum distance for masked pre-lifted QC-LDPC codes in $\xi^{QC}_{\matr{B}^{\uparrow 2}}(r)$  with parity-check matrix $\matr{H}=\matr{B}^{\uparrow 2 \upmapsto r}$ is increased to $\dmins \leq 34$.
 \end{itemize}\hfill $\Box$\end{example}

\begin{remark} {\em Note that very good irregular LDPC codes have been designed by optimizing their \emph{degree distribution} \cite{rsu01}. Masking applied to a pre-lifted base matrix $\bpre$ rather than the original base matrix $\matr{B}$ can give a code designer more flexibility in optimizing the degree distribution. In addition, optimizing degree distributions to improve performance in the waterfall region of the bit error rate (BER) curve often requires using many {low-degree variable nodes} \cite{rsu01}. In this case, pre-lifting can be used to good effect to mitigate the negative effect on minimum distance. This is particularly important for applications that require very low decoded BERs.}\end{remark}

%%%%%%%%%%%%%%%%%%%%%%%%%%%%%%%%%%%%%%%%%%%%%%%%%%%
\subsection{Protographs with repeated edges}\label{sec:multi}
%%%%%%%%%%%%%%%%%%%%%%%%%%%%%%%%%%%%%%%%%%%%%%%%%%%
In this section, we demonstrate the pre-lifting procedure applied to protographs with repeated edges. A great deal of effort has been devoted to {designing protograph-based code ensembles with desirable features} such as good iterative decoding thresholds and linear minimum distance growth. These protographs typically have repeated edges \cite{ddja09,ady07}.
\begin{example} The following $(3,4)$-regular example is taken from \cite{sv12}. Consider the $3\times 4$ base matrix
$$\matr{B} = \left[\begin{array}{cccc}
2 & 0 & 1 & 1\\
1 & 1 & 2 & 0\\
0 & 2 & 0 & 2
\end{array}\right].$$
Note that $\matr{B}$ has some entries greater than $1$; {consequently, when lifting the corresponding protograph to form a parity-check matrix $\matr{H}$, those entries $B_{i,j}$ are replaced by a summation of $B_{i,j}$ non-overlapping permutation matrices} (or circulant permutation matrices if desired).  It was shown in \cite{sv12} that the upper bound on minimum distance obtained for any circulant-based lifting of this base matrix is $\dmins \leq 32$. Recall that for the $3\times 4$ all-ones base matrix $\matr{B}$ we had $\dmins\leq 24$, so the upper bound on minimum distance is improved by including repeated edges. Indeed, the following circulant-based lifting of $\matr{B}$
$$\matr{H}\!=\!\matr{B}^{\upmapsto 46}\!\!=\!\!\left[\hspace{-1mm}\begin{array}{cccc} 
\matr{I}_1^{46}+\matr{I}_2^{46} &\matr{0} &\matr{I}_4^{46}&\matr{I}_8^{46}\\ 
\matr{I}_5^{46}& \matr{I}_9^{46}&\matr{I}_{10}^{46}+\matr{I}_{20}^{46}&\matr{0}\\ 
\matr{0}&\matr{I}_{25}^{46}+\matr{I}_{19}^{46}&\matr{0}&\matr{I}_7^{46}+\matr{I}_{14}^{46}
\end{array}\hspace{-1mm}\right]\hspace{-0.5mm}.$$
results in a $[184,47,32]$ code with girth $g=8$. 

Now consider the following pre-lifted base matrix
$$\matr{B}^{\uparrow 2} = \left[\begin{array}{cc|cc|cc|cc}
1 & 1 & 0 & 0 & 1 & 0 & 1 & 0\\
1 & 1 & 0 & 0 & 0 & 1 & 0 & 1\\\hline
1& 0 & 1 & 0 & 1 & 1 & 0 & 0\\
0 & 1 & 0 & 1 & 1 & 1 & 0 & 0\\\hline
0 & 0 & 1 & 1 & 0 & 0 & 1 & 1\\
0 & 0 & 1 & 1 & 0 & 0 & 1 & 1\\
\end{array}\right].$$
As a result of pre-lifting, the minimum distance of a code drawn from $\xi^{QC}_{\matr{B}^{\uparrow 2}}(r)$ with parity-check matrix $\matr{H}=\matr{B}^{\uparrow 2\upmapsto r}$ is bounded above by $\dmins \leq 108$, significantly larger than the upper bound $\dmins \leq 32$ obtained for codes drawn from $\xi^{QC}_{\matr{B}}(N)$ with parity-check matrix $\matr{H}=\matr{B}^{\upmapsto N}$.{\footnote{{If $B_{i,j}>1$, it is not required that the corresponding $B_{i,j}$ permutation matrices selected at the pre-lifting step are non-overlapping; however, this condition must be enforced at the second lifting step.}}} An example of a code with minimum distance exceeding the original bound is the null space of the following parity-check matrix
$$\matr{H}\!\!=\!\!\matr{B}^{\uparrow 2\upmapsto 46} \!\!= \!\!\left[\begin{array}{cc|cc|cc|cc}
\matr{I}_{1}^{46} & \matr{I}_{2}^{46} & 0 & 0 & \matr{I}_{4}^{46} & 0 & \matr{I}_{8}^{46} & 0\\
\matr{I}_{2}^{46} & \matr{I}_{0}^{46} & 0 & 0 & 0 & \matr{I}_{4}^{46} & 0 & \matr{I}_{8}^{46}\\\hline
\matr{I}_{5}^{46}& 0 & \matr{I}_{9}^{46} & 0 & \matr{I}_{10}^{46} & \matr{I}_{20}^{46} & 0 & 0\\
0 & \matr{I}_{5}^{46} & 0 & \matr{I}_{9}^{46} & \matr{I}_{20}^{46} & \matr{I}_{10}^{46} & 0 & 0\\\hline
0 & 0 & \matr{I}_{25}^{46} & \matr{I}_{19}^{46} & 0 & 0 & \matr{I}_{7}^{46} & \matr{I}_{14}^{46}\\
0 & 0 & \matr{I}_{19}^{46} & \matr{I}_{25}^{46} & 0 & 0 & \matr{I}_{14}^{46} & \matr{I}_{7}^{46}\\
\end{array}\right]\hspace{-0.5mm}.$$
This matrix defines a $[368,93,56]$ code with girth $g=8$, \emph{i.e.}, the minimum distance is significantly larger than the upper bound of $32$ for any {circulant-based $1$-cover} $\matr{H}=\matr{B}^{\upmapsto 2r}$. Note that the same design rules must be applied to protographs with repeated edges. In this example, the {circulant-block} matrix $\matr{H}_{1,1}$ in the upper-left corner ensures that the submatrices do not all commute and thus the upper bound can be exceeded. If, for example, we set 
$$\matr{H}_{1,1}=\left[\begin{array}{cc}
\matr{I}_{1}^{46} & \matr{I}_{2}^{46}\\
\matr{I}_{2}^{46} & \matr{I}_{1}^{46}
\end{array}\right],$$
then all the submatrices would commute and we would have $\dmins \leq 32$.
 \hfill $\Box$\end{example}

%%%%%%%%%%%%%%%%%%%%%%%%%%%%%%%%%%%%%%%%%%%%%%%%%%%
%%%%%%%%%%%%%%%%%%%%%%%%%%%%%%%%%%%%%%%%%%%%%%%%%%%
\section{{Code Design Rule 1}: Pre-lifting with \\circulant permutation matrices }\label{sec:liftcirculant}
%%%%%%%%%%%%%%%%%%%%%%%%%%%%%%%%%%%%%%%%%%%%%%%%%%%
%%%%%%%%%%%%%%%%%%%%%%%%%%%%%%%%%%%%%%%%%%%%%%%%%%%

In this section, we focus on Design Rule 1, where the pre-lifted protograph $\bpre$ is comprised of commuting submatrices. In particular, we consider the case when all of the permutation matrices $\matr{B}_{i,j}$ comprising $\bpre$ are circulant. By choosing the permutations at the pre-lifting step to be circulant, we can make use of their structure to eliminate many of the conditions that must otherwise be checked to achieve a desired girth $g$. Moreover, the conditions can be evaluated efficiently using modular arithmetic. 

%%%%%%%%%%%%%%%%%%%%%%%%%%%%%%%%%%%%%%%%%%%%%%%%%%%
\subsection{Girth conditions}\label{sec:circperm}
%%%%%%%%%%%%%%%%%%%%%%%%%%%%%%%%%%%%%%%%%%%%%%%%%%%
As noted previously, the technique given in \cite{smc11} can be used to generate a set of conditions on the permutation matrices comprising $\matr{H}=\matr{B}^{\uparrow N}$ that must be satisfied to guarantee a girth of at least $g$. Now consider a parity-check matrix $\matr{H}=\matr{B}^{\uparrow m \upmapsto r}$, where $N=mr$ and $m$ is the pre-lifting factor. By applying Lemma \ref{thm:girthlemma} to $\matr{H}$, we can eliminate many of the conditions that must be satisfied by a general matrix $\matr{H}=\matr{B}^{\uparrow mr}$ by checking if the corresponding products of the associated permutation matrices $\matr{B}_{i,j}$ comprising $\bpre$ have fixed columns. Choosing circulant permutation matrices at the pre-lifting step is advantageous for this purpose because we can quickly determine if a product of a number of circulant matrices has a fixed column using simple modular arithmetic (rather than costly matrix multiplication). This allows us to construct pre-lifted base matrices that reduce the number of conditions that must be satisfied in order to guarantee girth $g$ in Step 2 of the code design process.

\begin{example}
In this example, we focus on achieving $g\geq 8$ for a parity-check matrix in the form of (\ref{matrix3by4}), derived from a pre-lifted base matrix, but the same principles can be applied to a general protograph-based parity-check matrix derived from a pre-lifted base matrix for any desired girth. Suppose that $$\matr{P}=\diag(\matr{I}_{p_1}^r,\matr{I}_{p_2}^r,\ldots,\matr{I}_{p_m}^r)\cdot\tilde{\matr{I}}_{p}^m,$$ where $p\in[m]$ and $p_i\in[r]$, \emph{i.e.}, {circulant-block} matrix $\matr{P}$ is obtained by a double circulant-based lifting. (Similar definitions apply to $\matr{Q},\matr{R},\matr{S},\matr{T},$ and $\matr{U}$.) 

\begin{itemize}\item For pre-lifting factor $m=5$ and any pre-lifted base matrix $\base^{\upmapsto 5}$ obtained using circulant submatrices, the number of conditions (from the set (\ref{g8})) on the permutation matrices that comprise $\matr{H}=\base^{\upmapsto 5 \upmapsto r}$  that must be checked to guarantee $g\geq 8$ is in the range $[4,42]$. 
 \item Consider the following pre-lifted base matrix $\bpre$ in the form of (\ref{34preliftgeneral}) with $m=5$:
\begin{equation}
 \matr{B}^{\upmapsto 5}=\left[\begin{array}{cccc}
 \matr{I}_{0}^5 & \matr{I}_{0}^5 & \matr{I}_{0}^5 & \matr{I}_{0}^5\\
 \matr{I}_{0}^5 & \matr{I}_{0}^5 & \matr{I}_{1}^5 & \matr{I}_{1}^5\\
 \matr{I}_{0}^5 & \matr{I}_{0}^5 & \matr{I}_{2}^5 & \matr{I}_{4}^5
\end{array}\right].
\end{equation}
\noindent By choosing the permutation matrices given above at the pre-lifting step, we find that, in order to guarantee $g\geq 8$ in any resulting parity-check matrix $\matr{H}=\matr{B}^{\upmapsto 5 \upmapsto r}\in\xi_{\bpre}^{QC}(r)$, out of the $42$ original conditions given in (\ref{g8}), we only need to check that $\matr{P},\matr{Q},\matr{P}\matr{Q}^\tr$, and $\matr{R}\matr{T}^\tr$ do not have a fixed column. Equivalently, we must ensure $p_i \not\equiv 0\mod r$, $q_i \not\equiv 0\mod r$, $p_i+(r-q_i)\equiv p_i-q_i \not\equiv 0\mod r$, and $r_i-t_i \not\equiv 0\mod r$, $i=1,2,\ldots,5$. Since $\matr{S}$ and $\matr{U}$ are not involved in these four conditions, the values $s_i, u_i$, $i=1,\ldots,5$ can be chosen arbitrarily. 

\item In order to eliminate all the conditions given in (\ref{g8}), it is necessary to increase the pre-lifting factor to $m=9$. Then we find that it is possible to construct a pre-lifted base matrix $\bpre$ with circulant submatrices that has girth $8$. Consequently, by Corollary \ref{thm:girthcorollary}, any $\matr{H}\in\xi_{\bpre}^{QC}(r)$ satisfies $g\geq 8$, \emph{i.e.}, there are no conditions on the matrices $\matr{P}$, $\matr{Q}$, $\matr{R}$, $\matr{S}$, $\matr{T}$, and $\matr{U}$ that must be satisfied, so $p_i,q_i,r_i,s_i,t_i,$ and $u_i$, $i=1,2,\ldots,9$, can be chosen arbitrarily and we always obtain $g\geq 8$. The following pre-lifted base matrix is one such example:
\begin{equation}
  \matr{B}^{\upmapsto 9}=\left[\begin{array}{cccc}
 \matr{I}_{0}^9 & \matr{I}_{0}^9 & \matr{I}_{0}^9 & \matr{I}_{0}^9\\
 \matr{I}_{0}^9 & \matr{I}_{1}^9 & \matr{I}_{3}^9 & \matr{I}_{4}^9\\
 \matr{I}_{0}^9 & \matr{I}_{2}^9 & \matr{I}_{6}^9 & \matr{I}_{8}^9
\end{array}\right].
\end{equation}
\end{itemize}\end{example}\hfill $\Box$

%%%%%%%%%%%%%%%%%%%%%%%%%%%%%%%%%%%%%%%%%%%%%%%%%%%
\subsection{Minimum distance properties}
%%%%%%%%%%%%%%%%%%%%%%%%%%%%%%%%%%%%%%%%%%%%%%%%%%%

In this section, we construct a code using a circulant-based pre-lifting and show how its minimum distance is affected if we do not satisfy the overlapping column condition in Design Rule 1.

\begin{example}\label{ex:34mindist} Consider the pre-lifted base matrix $\base ^{\uparrow 2}$ given in (\ref{342cov}) with $m=2$ and the lifted parity-check matrix $\matr{H}\in\xi_{\bpre}^{QC}(r)$ given in (\ref{342cov2}). %\footnote{Note that, for $m=2$, we must use Design Rule 1 because all of the permutations are circulant.}  
Suppose that we set $p_1=p_2=1$, $q_1=q_2=7$, $r_1=r_2=10$, $s_1=s_2=11$, $t_1=t_2=13$, and $u_1=u_2=2$ with $r=49$. This parity-check matrix satisfies the conditions to achieve $g=10$. However, because the shift parameters in each {circulant-block matrix} are identical, this construction does not satisfy Design Rule 1. In fact, the conditions of Theorem \ref{thm:disttheorem} are satisfied, and the minimum distance is bounded above by {$d_\mathrm{min}\leq (n_c+1)!=24$}. This is in fact a $[392,100,24]$ QC code, \emph{i.e.}, the upper bound is achieved.

Suppose instead that we set $p_2=5$ and $u_2=4$, as in Example \ref{ex:34liftchoice2}, and denote the resulting code $\mathcal{C}_1$. Then the {circulant-block} permutation matrices $\matr{P}$ and $\matr{U}$ are comprised of two different circulant submatrices and, consequently, there exists a pair of strongly noncommutative matrices (\emph{e.g.}, $\matr{PQ}$ and $\matr{QP}$ for $r=49$), \emph{i.e.}, the conditions of Design Rule 1 are met. The minimum distance of $\mathcal{C}_1$ is increased to within the range {$32\leq d_\mathrm{min}\leq 56$} (determined using MAGMA) and $g=10$.
\end{example}\hfill $\Box$

%%%%%%%%%%%%%%%%%%%%%%%%%%%%%%%%%%%%%%%%%%%%%%%%%%%
\subsection{Simulation results}
%%%%%%%%%%%%%%%%%%%%%%%%%%%%%%%%%%%%%%%%%%%%%%%%%%%

\begin{figure}[h]
\begin{center}
\includegraphics[width=3.5in]{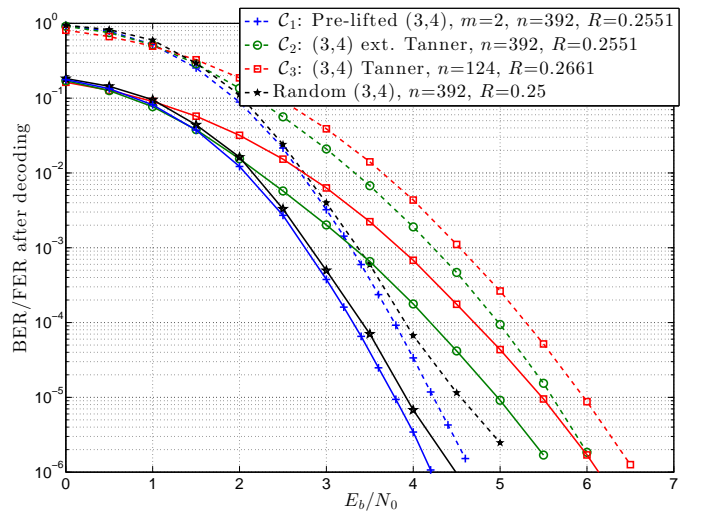}
\end{center}%\vspace{-3mm}
\caption{Simulated decoding performance in terms of BER (solid lines) and FER (dashed lines) for the pre-lifted $(3,4)$-regular QC-LDPC code $\mathcal{C}_1$ described in Example \ref{ex:34mindist}, the extended $(3,4)$-regular Tanner QC-LDPC code $\mathcal{C}_2$, the original Tanner code $\mathcal{C}_3$, and a randomly constructed $(3,4)$-regular code.}\label{fig:34_2cover}%\vspace{1mm}
\end{figure}

Computer simulations were performed assuming binary phase shift keyed (BPSK) modulation and an additive white Gaussian noise (AWGN) channel. The sum-product message passing decoder was allowed a maximum of $100$ iterations and employed a syndrome-check based stopping rule. In Fig. \ref{fig:34_2cover}, we plot the simulated decoding performance in terms of bit error rate (BER) and frame error rate (FER) for: the pre-lifted $(3,4)$-regular QC code $\mathcal{C}_1$ with $m=2$ from Example \ref{ex:34mindist}; the extended $(3,4)$-regular QC Tanner code with parity-check matrix defined in (\ref{tannercode}), denoted by  $\mathcal{C}_2$, %defined in Example \ref{ex:tanex}, 
where the circulant size is taken to be $N=98$ so that the code length and rate are the same as for code $\mathcal{C}_1$; and the original $(3,4)$-regular QC Tanner code with circulant size $N=31$, denoted by  $\mathcal{C}_3$. Both codes $\mathcal{C}_2$ and $\mathcal{C}_3$ achieve the upper bound {$d_\mathrm{min}=24$} and have $g=8$. We observe that the pre-lifted code $\mathcal{C}_1$ has significantly improved decoding performance, with a signal-to-noise ratio (SNR) gain of over $1$dB at a bit error rate of $10^{-5}$. Moreover, we also see from Fig. \ref{fig:34_2cover}, that the pre-lifted code outperforms a randomly constructed  $(3,4)$-regular code of the same length and slightly lower rate, particularly at high SNRs.

When sub-optimal decoding methods are employed, there are many factors in addition to the girth and minimum distance of a code that affect its performance (such as pseudocodewords, trapping sets, and  absorbing sets). Consequently, the improved simulated decoding performance of pre-lifted codes suggest pre-lifting may also improve these parameters.

%%%%%%%%%%%%%%%%%%%%%%%%%%%%%%%%%%%%%%%%%%%%%%%%%%%
%%%%%%%%%%%%%%%%%%%%%%%%%%%%%%%%%%%%%%%%%%%%%%%%%%%
\section{{Code Design Rule 2}: Pre-lifting with non-commuting permutation matrices}\label{sec:liftnoncommuting}
%%%%%%%%%%%%%%%%%%%%%%%%%%%%%%%%%%%%%%%%%%%%%%%%%%%
%%%%%%%%%%%%%%%%%%%%%%%%%%%%%%%%%%%%%%%%%%%%%%%%%%%

In this section, we construct a pre-lifted QC-LDPC code following Design Rule 2.

\begin{figure}[h]
\begin{center}
\includegraphics[width=3.5in]{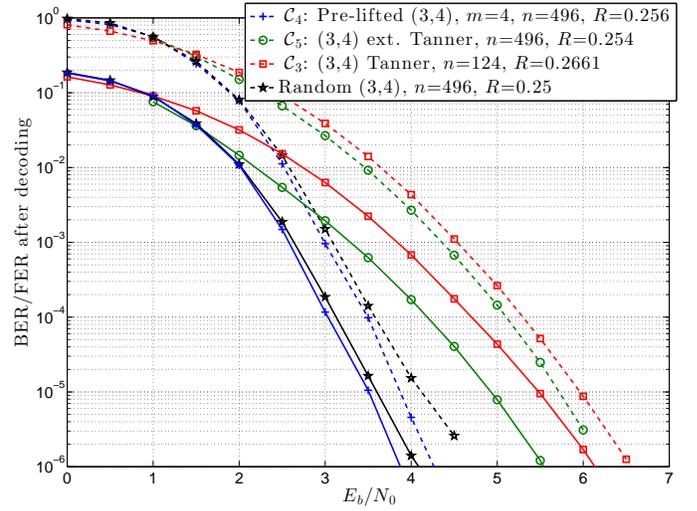}
\end{center}%\vspace{-2mm}
\caption{Simulated decoding performance  in terms of BER (solid lines) and FER (dashed lines) for the pre-lifted $(3,4)$-regular QC-LDPC code $\mathcal{C}_4$ described in Example \ref{ex:dr2}, the extended $(3,4)$-regular Tanner QC-LDPC code $\mathcal{C}_5$, the original Tanner code $\mathcal{C}_3$, and a randomly constructed $(3,4)$-regular code.}\label{fig:34tansim}%\vspace{-4mm}
\end{figure}

\begin{example}\label{ex:dr2} We construct a parity-check matrix $\matr{H}$ derived from a pre-lifted base matrix $\bpre$ defined in (\ref{34preliftgeneral}) with $m=4$. This matrix has the general form of (\ref{matrix3by4}), where $N=rm$ and 
\begin{align}
&\left[\hspace{-1mm}\begin{array}{ccc}
   \matr{P} &\matr{R}&\matr{T}\\
\matr{Q} &\matr{S}&\matr{U}
  \end{array}\right]=\left[\begin{array}{ccc}
   \matr{B}_{2,2}\otimes\matr{I}_{4}^r &\matr{B}_{2,3}\otimes\matr{I}_{12}^r &\matr{B}_{2,4}\otimes\matr{I}_{28}^r\\
\matr{B}_{3,2}\otimes\matr{I}_{24}^r &\matr{B}_{3,3}\otimes\matr{I}_{10}^r &\matr{B}_{3,4}\otimes\matr{I}_{13}^r
  \end{array}\right]   \nonumber \\&=\hspace{-1mm}\left[\begin{array}{cccc|cccc|cccc}
\matr{0}   & \hspace{-0.7mm}\matr{I}_4^r\hspace{-0.7mm} & \matr{0}   & \matr{0}   & \matr{0} & \matr{0} & \hspace{-0.7mm}\matr{I}_{{12}}^r\hspace{-0.7mm} & \matr{0}& \matr{0} & \hspace{-0.7mm}\matr{I}_{{28}}^r\hspace{-0.7mm} & \matr{0} & \matr{0}\\
\hspace{-0.7mm}\matr{I}_{4}^r\hspace{-0.7mm} & \matr{0}   & \matr{0}   & \matr{0}   & \matr{0} & \matr{0} & \matr{0} & \hspace{-0.7mm}\matr{I}_{{12}}^r\hspace{-0.7mm}& \matr{0} & \matr{0} & \hspace{-0.7mm}\matr{I}_{{28}}^r\hspace{-0.7mm} & \matr{0}\\
\matr{0}   & \matr{0}   & \matr{0}   & \hspace{-0.7mm}\matr{I}_{4}^r\hspace{-0.7mm} & \hspace{-0.7mm}\matr{I}_{{12}}^r\hspace{-0.7mm} & \matr{0} & \matr{0} & \matr{0}& \matr{0} & \matr{0} & \matr{0} & \hspace{-0.7mm}\matr{I}_{{28}}^r\hspace{-0.7mm}\\
\matr{0}   & \matr{0}   & \hspace{-0.7mm}\matr{I}_{4}^r\hspace{-0.7mm} & \matr{0}   & \matr{0} & \hspace{-0.7mm}\matr{I}_{{12}}^r\hspace{-0.7mm} & \matr{0} & \matr{0}& \hspace{-0.7mm}\matr{I}_{{28}}^r\hspace{-0.7mm} & \matr{0} & \matr{0} & \matr{0}\\\hline
\matr{0}   & \matr{0}   & \hspace{-0.7mm}\matr{I}_{{24}}^r\hspace{-0.7mm}   & \matr{0}   & \matr{0} & \hspace{-0.7mm}\matr{I}_{{10}}^r\hspace{-0.7mm} & \matr{0} & \matr{0}& \matr{0} & \matr{0} & \hspace{-0.7mm}\matr{I}_{{13}}^r\hspace{-0.7mm} & \matr{0}\\
\matr{0}   & \matr{0}   & \matr{0}   & \hspace{-0.7mm}\matr{I}_{{24}}^r\hspace{-0.7mm}   & \hspace{-0.7mm}\matr{I}_{{10}}^r\hspace{-0.7mm} & \matr{0} & \matr{0} & \matr{0}& \matr{0} & \matr{0} & \matr{0} & \hspace{-0.7mm}\matr{I}_{{13}}^r\hspace{-0.7mm}\\
\matr{0}   & \hspace{-0.7mm}\matr{I}_{{24}}^r\hspace{-0.7mm}   & \matr{0}   & \matr{0}   & \matr{0} & \matr{0} & \matr{0} & \hspace{-0.7mm}\matr{I}_{{10}}^r\hspace{-0.7mm}& \hspace{-0.7mm}\matr{I}_{{13}}^r\hspace{-0.7mm} & \matr{0} & \matr{0} & \matr{0}\\
\hspace{-0.7mm}\matr{I}_{{24}}^r\hspace{-0.7mm}   & \matr{0}   & \matr{0}   & \matr{0}   & \matr{0} & \matr{0} & \hspace{-0.7mm}\matr{I}_{{10}}^r\hspace{-0.7mm} & \matr{0}& \matr{0} & \hspace{-0.7mm}\matr{I}_{{13}}^r\hspace{-0.7mm} & \matr{0} & \matr{0}
\end{array}\right]\hspace{-1mm}.\nonumber
  \end{align}
~\\Note that the pre-lifting permutation matrices $\matr{B}_{2,2}$, $\matr{B}_{3,2}$, $\matr{B}_{2,3}$, $\matr{B}_{3,3}$, $\matr{B}_{2,4}$, and $\matr{B}_{3,4}$ have been chosen so that several pairs of permutation matrices are strongly noncommutative, yet $\matr{B}_{2,3}$, $\matr{B}_{2,4}$, and $\matr{B}_{3,4}$ are, in fact, circulant. The pre-lifting permutation matrices were chosen following the techniques presented in Section \ref{sec:constructingprelifted} in order to obtain large upper bounds on minimum distance. The shift parameters for each {circulant-block} permutation matrix were selected following the Tanner construction. Consequently, the pre-lifted Tanner graph associated with $\matr{H}$ can be considered as a $4$-fold graph cover of the original Tanner graph.

For $r=14$, we obtain a $[224,59,36]$ QC-LDPC code with $g=8$. As we increase $r$, the minimum distance generally improves, but it is difficult to verify the exact value using MAGMA as the code length increases. For $r=31$, we obtain a $[496,126]$ QC-LDPC code, denoted by $\mathcal{C}_4$, with $g=8$ and {$28\leq d_\mathrm{min}\leq 68$} (as in Example \ref{ex:34liftchoice1}, we conjecture that the minimum distance is, in fact, close to $68$).

In Fig. \ref{fig:34tansim}, we show the decoding performance of $\mathcal{C}_4$ and two $(3,4)$-regular QC Tanner codes: the extended $(3,4)$-regular QC Tanner code, denoted by $\mathcal{C}_5$, defined in (\ref{tannercode}), where the circulant size is taken to be $N=124$ so that the rate is approximately equal to that of $\mathcal{C}_4$ and the code lengths are equal; and the original $(3,4)$-regular QC Tanner code $\mathcal{C}_3$ with $N=31$. Again, we observe significantly improved decoding performance for the pre-lifted QC code. Moreover, we see that it performs slightly better than a randomly constructed  $(3,4)$-regular code of the same length and slightly lower rate, particularly at high SNRs.\end{example}\hfill $\Box$

Design Rule 2 is particularly useful because we can employ the theory presented in Section \ref{sec:constructingprelifted} to design a good pre-lifting matrix and use state-of-the-art QC codes, like the Tanner codes, to choose the circulants at Step 2 of the code design procedure. In the next section, we will see that large gains in decoding performance can be achieved by pre-lifting a `good' code. 

%%%%%%%%%%%%%%%%%%%%%%%%%%%%%%%%%%%%%%%%%%%%%%%%%%%
%%%%%%%%%%%%%%%%%%%%%%%%%%%%%%%%%%%%%%%%%%%%%%%%%%%
\section{Code design: Pre-lifting `good' codes}\label{sec:bocharova}
%%%%%%%%%%%%%%%%%%%%%%%%%%%%%%%%%%%%%%%%%%%%%%%%%%%
%%%%%%%%%%%%%%%%%%%%%%%%%%%%%%%%%%%%%%%%%%%%%%%%%%%

So far, we have used the Tanner code as a model, but Design Rule 2 can be applied to any array-based QC code. As a final example, we construct a nested family of QC-LDPC codes with design rates $R=1/4, 2/5, 1/2$, and $4/7$ using the pre-lifting technique. The model code we use is a $(3,7)$-regular QC-LDPC code with the following parity-check matrix \cite{bhj+12}:
\begin{equation}\label{bocharova}
\matr{H} = \matr{B}^{\upmapsto N}=\left[\begin{array}{ccccccc}
\matr{I}_{0}^N   & \matr{I}_{19}^N & \matr{I}_{13}^N & \matr{I}_{20}^N & \matr{I}_{4}^N & \matr{I}_{15}^N & \matr{I}_{56}^N\\
\matr{I}_{18}^N & \matr{I}_{9}^N   & \matr{I}_{0}^N   & \matr{I}_{47}^N & \matr{I}_{0}^N & \matr{I}_{18}^N & \matr{I}_{8}^N\\
\matr{I}_{14}^N & \matr{I}_{0}^N   & \matr{I}_{10}^N & \matr{I}_{13}^N & \matr{I}_{0}^N & \matr{I}_{0}^N   & \matr{I}_{7}^N
\end{array}\right],
\end{equation}
which can be obtained by a {one-step} circulant-based lifting of the $3\times 7$ all-ones base matrix $\matr{B}$. The circulants in this matrix were carefully selected using so-called `voltage graphs' in order to achieve a girth in the associated Tanner graph of $8$ with lifting factor $N=111$. Moreover, this code achieves the upper bound on minimum distance of $\dmins = 24$ for a direct circulant-based lifting of $\matr{B}$. {Note that this construction is `nested', in the sense that we can shorten this code} to be $(3,4)$-,  $(3,5)$-, $(3,6)$-, or $(3,7)$-regular by truncating (\ref{bocharova}) {from the right to have $4$, $5$, $6$, or $7$  (block) columns}, and, for the lifting factor $N=111$, each code will achieve the upper bound of $\dmins = 24$ for a direct circulant-based lifting of the corresponding (truncated) matrix $\matr{B}$. We will denote the code with $(3,K)$-regular parity-check matrix $\matr{H}$ and lifting factor $N$ by $\mathcal{C}_6(3,K,N)$.

In the following, we see that significantly improved decoding performance compared to this sequence of shortened codes can be obtained for each rate by carefully pre-lifting $\matr{B}$ and then using the same choice of circulants as in (\ref{bocharova}). In particular, we show empirically that the structural properties of the non-prelifted codes result in a `limiting performance' and almost identical error floors as the lifting factor $N$ increases, whereas the pre-lifted codes  exceed this performance with increasing $r$ as a result of their improved minimum distance.

Let $\matr{S}$ be the matrix of shift indices of $\matr{H}$, \emph{i.e.},
\begin{equation}
\matr{S} = \left[\begin{array}{ccccccc}
0    & 19 & 13 & 20 &   4 & 15 & 56\\
18  &   9 &   0 & 47 &   0 & 18 &    8\\
14  &   0 & 10 & 13 &   0 &   0 &    7
\end{array}\right].
\end{equation}
Then the parity-check matrix of the $(3,K)$-regular QC-LDPC code, $K \in \{4,5,6,7\}$, based on the pre-lifting is given by
\begin{equation}
\matr{H} = \left[\matr{H}_{i,j}\right]_{1\leq i \leq 3; 1 \leq j \leq K} = \left[ \matr{B}_{i,j} \otimes \matr{I}_{\matr{S}_{i,j}}^r \right]_{1\leq i \leq 3; 1 \leq j \leq K},
\end{equation} 
where the sub matrices $\matr{H}_{i,j}$ have size $mr\times mr$. We will denote the code with $(3,K)$-regular parity-check matrix $\matr{H}$ and lifting factor $r$ by $\mathcal{C}_7(3,K,m,r)$. Note that the girth of $\mathcal{C}_7(3,K,m,r)$ must be at least as large as the girth of  $\mathcal{C}_6(3,K,r)$. 

The pre-lifted matrix (\ref{bochprelift}) is obtained from $\matr{B}$ with pre-lifting factor $m=4$, where the permutation matrices were chosen following the techniques presented in Section \ref{sec:constructingprelifted} to give large upper bounds on the minimum distance and, adhering to Design Rule 2,  to ensure that at least one pair of submatrices $(\matr{B}_{i,j}, \matr{B}_{k,l})$, $(i,j)\neq (k,l)$, {is strongly noncommutative}. 
\bigformulatop{\value{equation}}
{ \begin{equation}\small
\base^{\uparrow 4}= \left[\begin{array}{cccc|cccc|cccc|cccc|cccc|cccc|cccc}\label{bochprelift}
1 & 0 & 0 & 0    & 1 & 0 & 0 & 0    & 1 & 0 & 0 & 0    & 1 & 0 & 0 & 0    & 1 & 0 & 0 & 0    & 1 & 0 & 0 & 0    & 1 & 0 & 0 & 0\\
0 & 1 & 0 & 0    & 0 & 1 & 0 & 0    & 0 & 1 & 0 & 0    & 0 & 1 & 0 & 0    & 0 & 1 & 0 & 0    & 0 & 1 & 0 & 0    & 0 & 1 & 0 & 0\\
0 & 0 & 1 & 0    & 0 & 0 & 1 & 0    & 0 & 0 & 1 & 0    & 0 & 0 & 1 & 0    & 0 & 0 & 1 & 0    & 0 & 0 & 1 & 0    & 0 & 0 & 1 & 0\\
0 & 0 & 0 & 1    & 0 & 0 & 0 & 1    & 0 & 0 & 0 & 1    & 0 & 0 & 0 & 1    & 0 & 0 & 0 & 1    & 0 & 0 & 0 & 1    & 0 & 0 & 0 & 1\\\hline

1 & 0 & 0 & 0    & 0 & 1 & 0 & 0    & 0 & 0 & 1 & 0    & 0 & 1 & 0 & 0    & 0 & 1 & 0 & 0    & 1 & 0 & 0 & 0    & 0 & 0 & 1 & 0\\
0 & 1 & 0 & 0    & 1 & 0 & 0 & 0    & 0 & 0 & 0 & 1    & 0 & 0 & 1 & 0    & 0 & 0 & 0 & 1    & 0 & 0 & 1 & 0    & 0 & 0 & 0 & 1\\
0 & 0 & 1 & 0    & 0 & 0 & 0 & 1    & 1 & 0 & 0 & 0    & 0 & 0 & 0 & 1    & 1 & 0 & 0 & 0    & 0 & 0 & 0 & 1    & 0 & 1 & 0 & 0\\
0 & 0 & 0 & 1    & 0 & 0 & 1 & 0    & 0 & 1 & 0 & 0    & 1 & 0 & 0 & 0    & 0 & 0 & 1 & 0    & 0 & 1 & 0 & 0    & 1 & 0 & 0 & 0\\\hline

1 & 0 & 0 & 0    & 0 & 0 & 1 & 0    & 0 & 1 & 0 & 0    & 0 & 0 & 1 & 0    & 0 & 0 & 1 & 0    & 0 & 1 & 0 & 0    & 1 & 0 & 0 & 0\\
0 & 1 & 0 & 0    & 0 & 0 & 0 & 1    & 1 & 0 & 0 & 0    & 0 & 0 & 0 & 1    & 1 & 0 & 0 & 0    & 0 & 0 & 1 & 0    & 0 & 0 & 1 & 0\\
0 & 0 & 1 & 0    & 0 & 1 & 0 & 0    & 0 & 0 & 0 & 1    & 1 & 0 & 0 & 0    & 0 & 0 & 0 & 1    & 1 & 0 & 0 & 0    & 0 & 0 & 0 & 1\\
0 & 0 & 0 & 1    & 1 & 0 & 0 & 0    & 0 & 0 & 1 & 0    & 0 & 1 & 0 & 0    & 0 & 1 & 0 & 0    & 0 & 0 & 0 & 1    & 0 & 1 & 0 & 0
\end{array}\right],
\end{equation}}\addtocounter{equation}{1}

\begin{figure}[h]
\begin{center}
\includegraphics[width=3.5in]{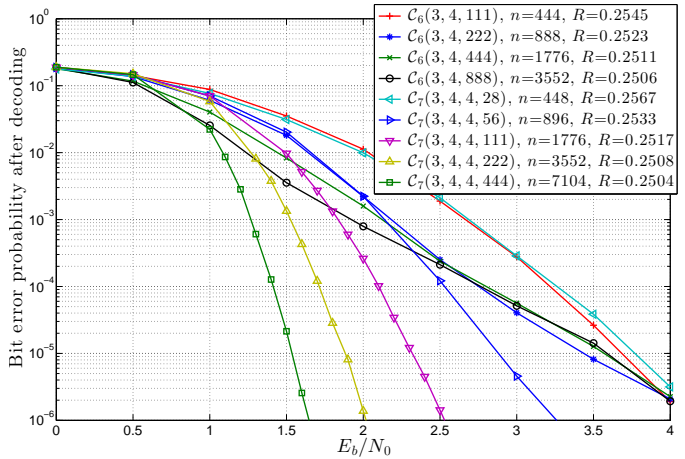}
\end{center}
\caption{Simulated decoding performance of several $(3,4)$-regular QC-LDPC codes $\mathcal{C}_6(3,4,N)$ and the pre-lifted $(3,4)$-regular QC-LDPC codes $\mathcal{C}_7(3,4,m,r)$ for a variety of lifting factors.}\label{fig:boch34}%
\end{figure}

Fig. \ref{fig:boch34} shows the decoding performance of $(3,4)$-regular QC-LDPC codes obtained for a variety of different lifting factors using circulant-based liftings of both the original base matrix $\matr{B}$ and the pre-lifted base matrix $\base ^{\uparrow 4}$. We observe for the $\mathcal{C}_6(3,4,N)$ codes that, as $N$ increases, the performance at low to moderate SNR improves (we observe an approximately $0.5 - 0.7$dB gain in the BER range $10^{-2}$ to $10^{-4}$ by increasing $N$ from $111$ to $222$, $444$, or $888$); however, at high SNRs the codes all suffer from an error floor (the BERs for $N=111, 222, 444$, and $888$ converge to approximately $2\times10^{-6}$ at an SNR of 4dB). The $\mathcal{C}_6(3,4,N)$ codes each have $g=10$.%Increasing $N$ further, we see that code $\mathcal{C}_6(3,4,888)$ displays more extreme performance,  having better performance at $1.5$dB, but an even higher error floor.

For the pre-lifted QC-LDPC codes $\mathcal{C}_7(3,4,4,r)$, we observe significantly improved decoding performance; in particular, we do not observe any error floors down to a BER of $10^{-6}$ for $r=111, 222$, and $444$, surpassing the `limiting performance' of the QC codes derived directly from $\matr{B}$. The $\mathcal{C}_7(3,4,4,r)$ codes each have $g=10$ for these lifting factors. We also include the decoding performance for smaller lifting factors $r=28$ and $r=56$, even though the circulants were optimized in \cite{bhj+12} for $N=111$. Consequently, these codes have reduced girth $g=6$; however, we see that, for $r=28$, the pre-lifted code $\mathcal{C}_7(3,4,4,28)$ has approximately the same decoding performance as $\mathcal{C}_6(3,4,111)$, illustrating that (\ref{bochprelift}) represents a good choice for $\base ^{\uparrow 4}$, and for $r=56$ we see improved performance in the high SNR region compared to any of the one-step liftings (even those with larger block lengths). Finally, we note that the performance of the pre-lifted code $\mathcal{C}_7(3,4,4,444)$ is only about $0.4$dB from the iterative decoding threshold $\gamma_{\mathrm{iter}}=1.2758$dB of the $(3,4)$-regular protograph-based ensemble $\xi_\matr{B}(N)$ at a BER of $10^{-6}$, and we would expect this gap to decrease as we increase $r$.\footnote{Iterative decoding thresholds for the AWGN channel were estimated using the reciprocal channel approximation (RCA) technique \cite{chu00}.} Our results indicate that similar performance is unlikely to be realized for the $\mathcal{C}_6(3,4,N)$ codes, even by letting $N$ become very large, since these codes have limited minimum distance.

\begin{figure}[h!]
\begin{center}
\includegraphics[width=3.5in]{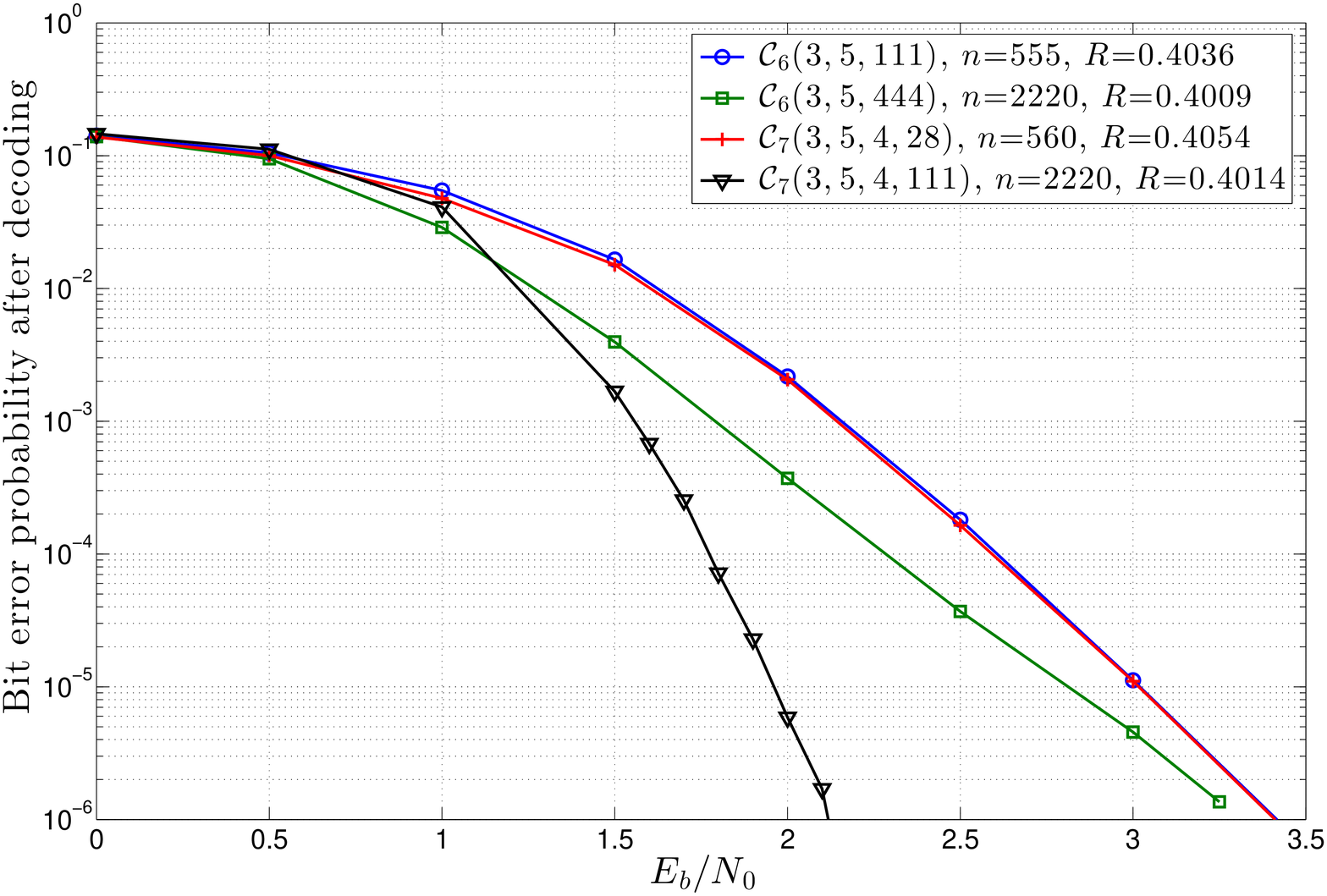}
\end{center}%\vspace{-10mm}
\caption{Simulated decoding performance of several $(3,5)$-regular QC-LDPC codes $\mathcal{C}_6(3,5,N)$ and the pre-lifted $(3,5)$-regular QC-LDPC codes $\mathcal{C}_7(3,5,4,r)$ for a variety of lifting factors.}\label{fig:boch35}
\end{figure}

\begin{figure}[h!]
\begin{center}
\includegraphics[width=3.5in]{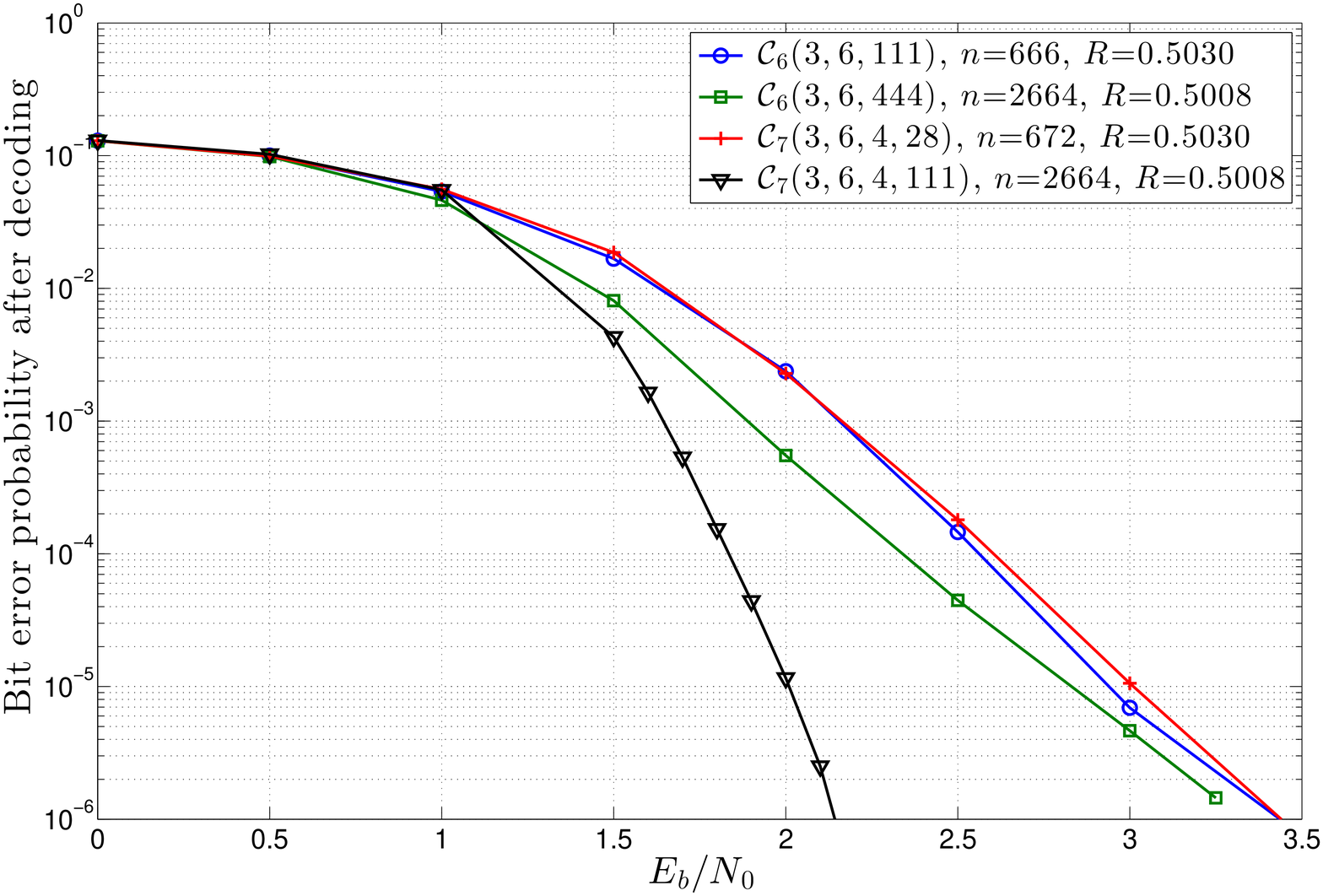}
\end{center}\vspace{-4mm}
\caption{Simulated decoding performance of several $(3,6)$-regular QC-LDPC codes $\mathcal{C}_6(3,6,N)$ and the pre-lifted $(3,6)$-regular QC-LDPC codes $\mathcal{C}_7(3,6,4,r)$ for a variety of lifting factors.}\label{fig:boch36}
\end{figure}

Figures \ref{fig:boch35}, \ref{fig:boch36}, and \ref{fig:boch37} show the decoding performance of the higher rate $(3,5)$-, $(3,6)$-, and $(3,7)$-regular QC-LDPC codes, respectively, obtained for a variety of different lifting factors using circulant-based liftings of both the original base matrix $\matr{B}$ and the pre-lifted base matrix $\base ^{\uparrow 4}$. We again see that the performance of the $4$-covers with small lifting factor $r=28$ is approximately equal to the performance of the original code, indicating that (\ref{bochprelift}) represents a good choice for $\base ^{\uparrow 4}$ in each case. For each code rate, we see that the pre-lifted code $\mathcal{C}_7(J,K,4,111)$ outperforms the once-lifted code $\mathcal{C}_6(J,K,444)$ of the same length and approximately the same rate. Moreover, we expect that this gap will increase as we further increase the lifting factors. Finally, we note that the performance of the code $\mathcal{C}_7(3,6,4,111)$ in Fig.~\ref{fig:boch36} is only slightly worse (about $0.1-0.2$dB) than the $[2304,1152]$ WiMAX code \cite{wimax}, despite the fact that the construction described above involved only an easy search for a good pre-lifting matrix and then simply adopted \eqref{bocharova} for the second lifting step.

{\begin{figure}[h!]
\begin{center}
\includegraphics[width=3.5in]{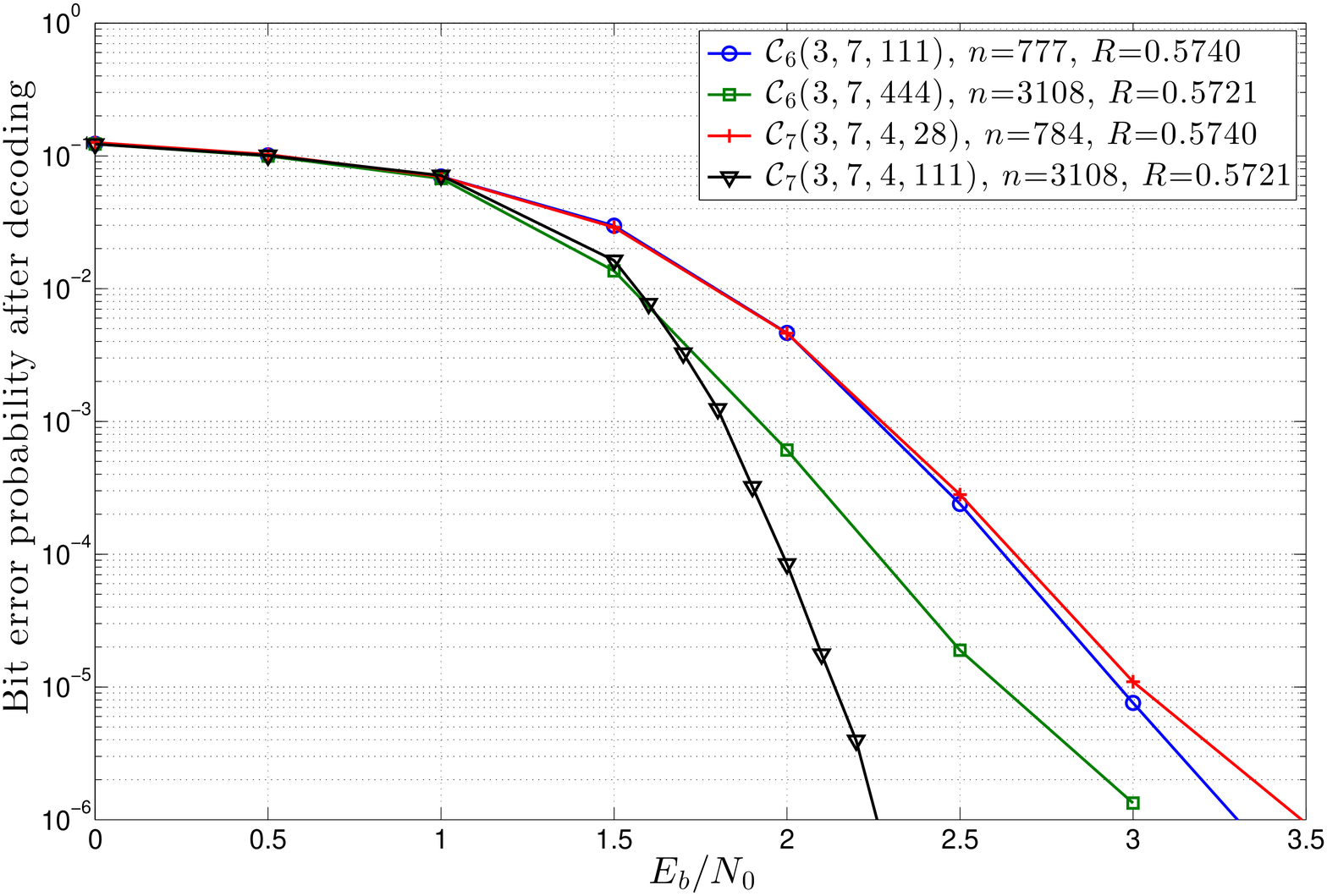}
\end{center}\vspace{-4mm}
\caption{Simulated decoding performance of several $(3,7)$-regular QC-LDPC codes $\mathcal{C}_6(3,7,N)$ and the pre-lifted $(3,7)$-regular QC-LDPC codes $\mathcal{C}_7(3,7,4,r)$ for a variety of lifting factors.}\label{fig:boch37}
\end{figure}

We have seen that the `limiting performance' of one-step circulant liftings of a code can be exceeded by a pre-lifted code. This result indicates that to design QC-LDPC codes with larger block lengths, it may be better to pre-lift the base matrix rather than increase the circulant size, since QC-LDPC codes based on pre-lifted protographs have improved minimum distance and large girth. We attributed the improved decoding performance reported in Sections \ref{sec:liftcirculant}-\ref{sec:bocharova} to these parameters; however, when sub-optimal iterative decoding methods are employed, there are other parameters in addition to girth and minimum distance that affect code performance (such as pseudocodewords, trapping sets, and  absorbing sets). Consequently, the improved simulated decoding performance of pre-lifted codes suggests that pre-lifting may also improve these parameters.

%%%%%%%%%%%%%%%%%%%%%%%%%%%%%%%%%%%%%%%%%%%%%%%%%%%
\section{Concluding Remarks}\label{sec:conclusions}
%%%%%%%%%%%%%%%%%%%%%%%%%%%%%%%%%%%%%%%%%%%%%%%%%%%

In this paper, we presented new results on QC-LDPC codes that are constructed using a two-step lifting procedure based on a protograph, and, by implementing this method instead of the usual one-step procedure, we were able to show improved minimum distance and girth properties. We also presented two design rules to construct QC-LDPC codes based on pre-lifting: one uses only commuting pairs of permutation matrices at the first (pre-lifting) stage, while the other includes some strongly noncommutative pairs of permutation matrices. For both design rules, we obtained an increase in minimum distance compared to a one-step circulant-based lifting and improved performance was verified by simulation. Finally, we showed that the pre-lifting technique can also be applied to any `good' QC-LDPC code existing in the literature and results in a new QC-LDPC code with improved minimum distance, girth, and decoding performance.

\appendices
\section{} \label{sec:applift}
The matrix $\matr{H} = \matr{B}^{\uparrow N}$ consisting of an $n_c\times n_v$ array of permutation matrices $\matr{Q}_{i,j}$, $i\in \{1,2,\ldots ,n_c\}$,  $ j\in \{1,2,\ldots ,n_v\}$, can be transformed by column operations into
\begin{align} \label{matrix:general2}
{\small\begin{bmatrix} 
\matr{I}_0^N                                 \!\! &\!\!     \matr{I}_0^N                                        &\!\!\cdots&\!\! \matr{I}_0^N     \\
\matr{Q}_{2,1}\matr{Q}_{1,1}^\tr      \!\! &\!\!\matr{Q}_{2,2}\matr{Q}_{1,2}^\tr                 &\!\!\cdots&\!\!\matr{Q}_{2,n_v}\matr{Q}_{1,n_v}^\tr\\
%\matr{Q}_{20}\matr{Q}_{00} ^\tr     \!\! &\!\!   \matr{Q}_{21}\matr{Q}_{01}^\tr              &\!\!\cdots&\!\!\matr{Q}_{2,n-1}\matr{Q}_{0,n-1}^\tr\\ 
\vdots                        \!\! &\!\!            \vdots                     &&\!\!\vdots\\
\matr{Q}_{n_c,1}\matr{Q}_{1,1}^\tr \!\! & \!\! \matr{Q}_{n_c,2}\matr{Q}_{1,2}^\tr         &\!\!\cdots&\!\!\matr{Q}_{n_c,n_v}\matr{Q}_{1,n_v}^\tr
\end{bmatrix}},
\end{align}
followed by  row operations to  transform it into
\begin{align} \label{matrix:general3}
{\small\hspace{-1mm}\left[\hspace{-1mm}
\begin{array}{cccc}
\matr{I}_0^N                                 \!\! &\!\!     \matr{I}_0^N                                        &\!\!\cdots&\!\! \matr{I}_0^N     \\
 \matr{I}_0^N      \!\! &\!\!\left(\matr{Q}_{2,1}\matr{Q}_{1,1}^\tr\right)^\tr \matr{Q}_{2,2}\matr{Q}_{1,2}^\tr                 &\!\!\cdots&\!\!\left(\matr{Q}_{2,1}\matr{Q}_{1,1}^\tr\right)^\tr\matr{Q}_{2,n_v}\matr{Q}_{1,n_v}^\tr\\
%\matr{Q}_{20}\matr{Q}_{00} ^\tr     \!\! &\!\!   \matr{Q}_{21}\matr{Q}_{01}^\tr              &\!\!\cdots&\!\!\matr{Q}_{2,n-1}\matr{Q}_{0,n-1}^\tr\\ 
\vdots                        \!\! &\!\!            \vdots                     &&\!\!\vdots\\
\hspace{0mm}\matr{I}_0^N\hspace{0mm} & \hspace{-1mm}\!\! \hspace{-1mm}\left(\matr{Q}_{n_c,1}\matr{Q}_{1,1}^\tr\right)^\tr   \matr{Q}_{n_c,2}\matr{Q}_{1,2}^\tr\hspace{-1mm}        &\hspace{-1mm}\!\!\cdots\hspace{-1mm}&\!\! \hspace{-1mm}\left(\matr{Q}_{n_c,1}\matr{Q}_{1,1}^\tr\right)^\tr \matr{Q}_{n_c,n_v}\matr{Q}_{1,n_v}^\tr\hspace{-1mm}
\end{array}
\hspace{-1mm}\right]\hspace{-1mm}.}
\end{align}
\noindent The operations described above do not affect the girth of the Tanner graph or  the minimum distance of the code because we have simply reordered the rows and columns. If all permutation matrices $\matr{Q}_{i,j}$ are circulant (or {circulant-block}) then the products of such matrices as  in \eqref{matrix:general3} must also be circulant (resp. {circulant-block}) by Property 2 of circulant permutation matrices.
\section{} \label{sec:appa}
{\it Proof of Lemma \ref{thm:girthlemma}}.
{Suppose $\matr{P}=\diag(\matr{I}_{p_1}^r,\matr{I}_{p_2}^r,\ldots,\matr{I}_{p_m}^r)\cdot\tilde{\matr{B}}_P$ and
$\matr{Q}=\diag(\matr{I}_{q_1}^r,\matr{I}_{q_2}^r,\ldots,\matr{I}_{q_m}^r)\cdot\tilde{\matr{B}}_Q$, so that}
\begin{equation}
 \matr{PQ}=\diag(\matr{I}_{p_1}^r,\matr{I}_{p_2}^r,\ldots,\matr{I}_{p_m}^r)\tilde{\matr{B}}_P\cdot\diag(\matr{I}_{q_1}^r,\matr{I}_{q_2}^r,\ldots,\matr{I}_{q_m}^r)\tilde{\matr{B}}_Q,
\end{equation}
where $p_i,q_i\in[r]$, $i\in [m]$,  $\matr{B}_P$ and $\matr{B}_Q$ are $m\times m$ permutation matrices, and $\tilde{\matr{B}}_{P}= \matr{B}_{P}\otimes \matr{I}_0^r$ (as defined in Section \ref{sec:prelifting}-A). Then
\begin{align*}
 \matr{PQ}=&\\&\diag(\matr{I}_{p_1}^r, \ldots,\matr{I}_{p_m}^r)\diag(\matr{I}_{q_{\sigma(1)}}^r,\ldots,\matr{I}_{q_{\sigma(m)}}^r)\cdot\tilde{\matr{B}}_P\tilde{\matr{B}}_Q,
=\\&\diag(\matr{I}^r_{(p_1+q_{\sigma(1)})\hspace{-2mm}\mod r},\ldots,\matr{I}^r_{(p_m+q_{\sigma(m)})\hspace{-2mm}\mod r})\cdot\tilde{\matr{B}}_P\tilde{\matr{B}}_Q,
\end{align*}
where $\sigma$ is the permutation associated with permutation matrix $\matr{B}_P$. Note also that by the distributive law of the Kronecker product
\begin{align*}
\tilde{\matr{B}}_P\tilde{\matr{B}}_Q =\left(\matr{B}_P\otimes \matr{I}_0^r\right)\cdot\left(\matr{B}_Q\otimes \matr{I}_0^r\right)=\matr{B}_P\matr{B}_Q\otimes \matr{I}_0^r\matr{I}_0^r.   %\vspace{-2mm}  
                                                                                                                                                                                                                                 \end{align*}
Suppose $\matr{B}_{P}\matr{B}_{Q}$ does not have a fixed column, \emph{i.e.}, $(\matr{B}_{P}\matr{B}_{Q})_{k,k}=0$, $k=1,2,\ldots,m$. Then $(\tilde{\matr{B}}_P\tilde{\matr{B}}_Q)_{i,j}=0$ for $(i,j)\in \mathcal{S}$, where
\begin{equation}
\mathcal{S}=\mathop{\cup}_{k=1}^m \left(\mathcal{S}_k\times\mathcal{S}_k\right)=\mathop{\cup}_{k=1}^m \mathcal{S}_k^2,
\end{equation}
$\mathcal{S}_k=\{(k-1)r+1,(k-1)r+2,\ldots,kr\}$, and $\mathcal{S}_k\times\mathcal{S}_k$ denotes the Cartesian product of two sets.
Now, 
\begin{align}
&( \matr{PQ})_{i,i}  \nonumber\\&=\sum_{j=1}^{mr} \left(\diag(\matr{I}^r_{(p_1+q_{\sigma(1)})},\ldots,\matr{I}^r_{(p_m+q_{\sigma(m)})})\right)_{i,j}(\tilde{\matr{B}}_P\tilde{\matr{B}}_Q)_{j,i},\label{prodfixedcol}
\end{align}
for $i \in\{1,2,\ldots,mr\}$. Suppose $i \in \mathcal{S}_k$; then, it follows from the structure of a block diagonal permutation matrix that the only non-zero symbol $(\diag(\matr{I}^r_{(p_1+q_{\sigma(1)})},\matr{I}^r_{(p_2+q_{\sigma(2)})},\ldots,\matr{I}^r_{(p_m+q_{\sigma(m)})}))_{i,j}$ occurs when $j \in \mathcal{S}_k$. Then $(j,i) \in \mathcal{S}$, which implies $(\tilde{\matr{B}}_P\tilde{\matr{B}}_Q)_{j,i}=0$, and thus $(\matr{PQ})_{i,i} = 0$ follows from (\ref{prodfixedcol}).\hfill$\Box$

 \section{} \label{sec:appb}
%\newtheorem{distlemma}[mddisttheorem]{\bf Lemma}
%\begin{distlemma}
\emph{Proof of Lemma \ref{thm:distlemma}}.
($\Rightarrow$) Suppose that $\matr{PQ}=\matr{QP}$. Then 
\begin{align*}
\matr{PQ}&-\matr{QP}\\&=(\matr{B}_P\otimes\matr{I}_{p_1}^r)(\matr{B}_Q\otimes\matr{I}_{q_1}^r)-(\matr{B}_Q\otimes\matr{I}_{q_1}^r)(\matr{B}_P\otimes\matr{I}_{p_1}^r)\\&=\matr{B}_P\matr{B}_Q\otimes\matr{I}_{p_1}^r\matr{I}_{q_1}^r-\matr{B}_Q\matr{B}_P\otimes\matr{I}_{q_1}^r\matr{I}_{p_1}^r\\&=(\matr{B}_P\matr{B}_Q-\matr{B}_Q\matr{B}_P)\otimes\matr{I}^r_{(p_1+q_1)\hspace{-1mm}\mod r}=\matr{0},
\end{align*}
which implies that $\matr{B}_P\matr{B}_Q=\matr{B}_Q\matr{B}_P$ because $\matr{I}^r_{(p_1+q_1)\hspace{-1mm}\mod r}\neq\matr{0}$.
\vspace{1mm}

 ($\Leftarrow$) If $\matr{B}_P\matr{B}_Q=\matr{B}_Q\matr{B}_P$, then  \begin{align*}\matr{PQ}&=\matr{B}_P\matr{B}_Q\otimes\matr{I}^r_{(p_1+q_1)\hspace{-1mm}\mod r}\\&=\matr{B}_Q\matr{B}_P\otimes\matr{I}^r_{(q_1+p_1)\hspace{-1mm}\mod r}=\matr{QP}.\end{align*}\hfill$\Box$

 \section{} \label{sec:appc}
\begin{example}\label{ex:md} 
Consider the $2\times 3$ all-ones base matrix $\matr{B}$. Suppose that $\matr{B}$ is lifted twice to form the QC-LDPC parity-check matrix 
$$ \matr{H}=\hpre=\left[\begin{array}{ccc}
 \matr{H}_{1,1} &  \matr{H}_{1,2} & \matr{H}_{1,3}\\
 \matr{H}_{2,1} & \matr{H}_{2,2} &  \matr{H}_{2,3}
\end{array}\right]_{2mr\times 3mr},$$
where the {circulant-block} permutation matrices $\matr{H}_{i,j}$ are all non-overlapping.
After row and column permutations, $\matr{H}$ can be re-written as
$$ \matr{H}=\left[\begin{array}{ccc}
 \matr{I}_{0}^{mr} & \matr{I}_{0}^{mr} & \matr{I}_{0}^{mr}\\
 \matr{I}_{0}^{mr} & \matr{P} & \matr{Q}
\end{array}\right],$$
where $\matr{P}$ and $\matr{Q}$ are {circulant-block}. (This re-writing of $\matr{H}$ is not necessary, but it simplifies the following arguments.) Now consider the $3mr$-tuple
\begin{equation*}\label{mdcodeword}
\matr{c}^\tr = \left[\begin{array}{ccc}
(\matr{P}+\matr{Q})\matr{x};&( \matr{I}_{0}^{mr}+\matr{Q})\matr{x};&( \matr{I}_{0}^{mr}+\matr{P})\matr{x}
\end{array}\right],
\end{equation*}
where $\matr{x}$ is a arbitrary weight one column vector %, say $\matr{x}=[\>1\> 0 \>\cdots \>0\>]^\tr$, 
and ``;'' is used to denote stacking of column vectors. Since the permutation matrices comprising $\matr{H}$ are non-overlapping, the Hamming weight of $\matr{c}$ is  ${\rm wt}(\matr{c}) = (2+1)! = 6$. Then
\begin{align}
\matr{s}^\tr &=  \matr{H}\cdot\matr{c}^\tr \nonumber
\left[\begin{array}{ccc}
(\matr{P}+\matr{Q}+\matr{I}_{0}^{mr}+\matr{Q}+\matr{I}_{0}^{mr}+\matr{P})\matr{x}\\
(\matr{P}+\matr{Q}+\matr{P}+\matr{PQ}+\matr{Q}+\matr{QP})\matr{x}
\end{array}\right] \nonumber\\ &=
\left[\begin{array}{ccc}
\matr{0}\\
(\matr{PQ}+\matr{QP})\matr{x}
\end{array}\right],
\end{align}
and $\matr{c}$ is a codeword if and only if $ (\matr{PQ}+\matr{QP})\matr{x}=\matr{0}$. Consequently, if $\matr{PQ}$ and $\matr{QP}$ have an overlapping column, \emph{i.e.,} if $\matr{P}$ and $\matr{Q}$ are not strongly noncommutative, then there exists an $\matr{x}$ such that $\matr{c}$ is a codeword. If $\matr{P}$ and $\matr{Q}$ are strongly noncommutative, then $ (\matr{PQ}+\matr{QP})\matr{x}\neq\matr{0}$ and $\matr{c}$ corresponds to a $(6,f)$ \emph{near-codeword}, where the Hamming weight of the syndrome vector ${\rm wt}(\matr{s}) = f$ denotes the number of unsatisfied parity-check equations.
%\end{distremark}
\end{example}\hfill $\Box$
\end{document}